\documentclass[%
 reprint,
 superscriptaddress,
 nofootinbib,
 amsmath,amssymb,
 prc,
 floatfix
]{revtex4-2}

\usepackage{graphicx}
\usepackage{dcolumn}
\usepackage{bm}
\usepackage{xcolor}
\usepackage{wasysym}
\usepackage{physics}
\usepackage[pdftex,colorlinks,linkcolor = blue,urlcolor  = blue,citecolor = blue]{hyperref}

\begin{document}

\preprint{APS/123-QED}

\title{Nuclear level densities and $\gamma-$ray strength functions in $^{120,124}$Sn isotopes: impact of Porter-Thomas fluctuations}

\author{M.~Markova}
\email{maria.markova@fys.uio.no}
\affiliation{Department of Physics, University of Oslo, N-0316 Oslo, Norway}


\author{A.~C.~Larsen}
\email{a.c.larsen@fys.uio.no}
\affiliation{Department of Physics, University of Oslo, N-0316 Oslo, Norway}

\author{P.~von Neumann-Cosel}%
\affiliation{%
 Institut f\"{u}r Kernphysik, Technische Universit\"{a}t Darmstadt, D-64289 Darmstadt, Germany
}%

\author{S.~Bassauer}%
\affiliation{%
 Institut f\"{u}r Kernphysik, Technische Universit\"{a}t Darmstadt, D-64289 Darmstadt, Germany
}%

\author{A.~G\"{o}rgen}
\affiliation{Department of Physics, University of Oslo, N-0316 Oslo, Norway}

\author{M.~Guttormsen}
\affiliation{Department of Physics, University of Oslo, N-0316 Oslo, Norway}

\author{F.~L.~Bello Garrote}
\affiliation{Department of Physics, University of Oslo, N-0316 Oslo, Norway}

\author{H. C.~Berg}
\affiliation{Department of Physics, University of Oslo, N-0316 Oslo, Norway}

\author{M.~M.~Bj{\o}r{\o}en}
\affiliation{Department of Physics, University of Oslo, N-0316 Oslo, Norway}

\author{T.~K.~Eriksen}
\affiliation{Department of Physics, University of Oslo, N-0316 Oslo, Norway}

\author{D.~Gjestvang}
\affiliation{Department of Physics, University of Oslo, N-0316 Oslo, Norway}

\author{J.~Isaak}%
\affiliation{%
 Institut f\"{u}r Kernphysik, Technische Universit\"{a}t Darmstadt, D-64289 Darmstadt, Germany
}%

\author{M.~Mbabane}
\affiliation{Department of Physics, University of Oslo, N-0316 Oslo, Norway}

\author{W.~Paulsen}
\affiliation{Department of Physics, University of Oslo, N-0316 Oslo, Norway}

\author{L.~G.~Pedersen}
\affiliation{Department of Physics, University of Oslo, N-0316 Oslo, Norway}

\author{N.~I.~J.~Pettersen}
\affiliation{Department of Physics, University of Oslo, N-0316 Oslo, Norway}

\author{A.~Richter}
\affiliation{%
 Institut f\"{u}r Kernphysik, Technische Universit\"{a}t Darmstadt, D-64289 Darmstadt, Germany
}%

\author{E.~Sahin}
\affiliation{Department of Physics, University of Oslo, N-0316 Oslo, Norway}

\author{P.~Scholz}
\affiliation{Institut f\"{u}r Kernphysik, Universit\"{a}t zu K\"{o}ln, D-50937 K\"{o}ln, Germany}
\affiliation{Department of Physics, University of Notre Dame, Indiana 46556-5670, USA}

\author{S.~Siem}
\affiliation{Department of Physics, University of Oslo, N-0316 Oslo, Norway}

\author{G.~M.~Tveten}
\affiliation{Department of Physics, University of Oslo, N-0316 Oslo, Norway}

\author{V.~M.~Valsdottir}
\affiliation{Department of Physics, University of Oslo, N-0316 Oslo, Norway}

\author{M.~Wiedeking}
\affiliation{Deptartment of Subatomic Physics to SSC Laboratory, iThemba LABS, Somerset West 7129, South Africa}
\affiliation{School of Physics, University of the Witwatersrand, Johannesburg 2050, South Africa}

\date{\today}

\begin{abstract}

 Nuclear level densities (NLDs) and $\gamma$-ray strength functions (GSFs) of $^{120,124}$Sn have been extracted with the Oslo method from  proton-$\gamma$ coincidences in the ($p,p^{\prime}\gamma)$ reaction. 
 The functional forms of the GSFs and NLDs have been further constrained with the Shape method by studying primary $\gamma$-transitions to the ground and first excited states. 
 The NLDs demonstrate good agreement with the NLDs of $^{116,118,122}$Sn isotopes measured previously. 
 Moreover, the extracted partial NLD of 1$^{-}$ levels in $^{124}$Sn is shown to be in fair agreement with those deduced from spectra of relativistic Coulomb excitation in forward-angle inelastic proton scattering. 
 
 The experimental NLDs have been applied to estimate the magnitude of the Porter-Thomas (PT) fluctuations. 
 Within the PT fluctuations,  we conclude that the GSFs for both isotopes can be considered to be independent of initial and final excitation energies, in accordance with the generalized Brink-Axel hypothesis. 
 Particularly large fluctuations observed in the Shape-method GSFs present a considerable contribution to the uncertainty of the method, and may be one of the reasons for deviations from the Oslo-method strength at low $\gamma$-ray energies and low values of the NLD (below $\approx1\cdot10^{3}-2\cdot10^{3}$ MeV$^{-1}$).

\end{abstract}

\maketitle
\section{\label{sec 1: introduction}Introduction}

Numerous experimental and theoretical efforts  have been dedicated to the study of $\gamma-$decay processes in atomic nuclei. 
The decay properties of excited nuclei are not only pivotal for the basic nuclear physics research, but also are the core ingredients for large-scale calculations of nucleosynthesis and element abundances in the universe \cite{Goriely1998, Arnould2007}. 

While gradually moving from the lowest to higher excitation energies of a nucleus, the spacing between individual excited states becomes smaller, and the sensitivity of experimental techniques might be no longer sufficient to resolve them separately. Here, the nucleus enters the quasi-continuum regime and the concept of the nuclear level density (NLD), \textit{i.e.} the number of nuclear states per excitation energy unit, becomes an indispensable tool for a statistical description of nuclei. 
By analogy, the $\gamma$-ray strength function (GSF), or the average, reduced $\gamma$-transition probability, becomes more suitable to describe the numerous $\gamma$-transitions.
The statistical model as formulated by Hauser and Feshbach \cite{Hauser1952} with ingredients such as the NLD and GSF, provides the main framework for modelling nuclear reactions and calculating their cross-sections for astrophysical purposes (\textit{e.g.} \cite{Arnould2007}), the design of nuclear reactors \cite{Chadwick2011}, and the transmutation of nuclear waste \cite{Salvatore2011}.

Among all experimental techniques used for the extraction of GSFs \cite{Goriely2019a}, the Oslo method has been widely used to obtain the dipole strength below the neutron threshold by studying the $\gamma$-decay of residual nuclei formed in light-ion induced reactions \cite{Guttormsen1996, Guttormsen1987, Schiller2000}. 
The main advantage of the method is a simultaneous extraction of the NLD and GSF from  primary $\gamma-$decay spectra at excitation energies below the neutron separation energy $S_n$. 
The GSFs for many nuclei obtained by employing different experimental techniques have previously been reported to provide a rather good agreement in absolute values and/or general shapes with the Oslo method strengths \cite{Wiedeking2012, Bassauer2016, Martin2017}. A few cases of large discrepancies have also been reported (\textit{e.g.} the comparison of the Oslo and ($\gamma,\gamma^{\prime})$ data for $^{89}$Y and $^{139}$La presented in Ref. \cite{Goriely2019a}).

A large fraction of theoretical and experimental techniques focusing on calculating or measuring the GSF, including the Oslo method, rely on the validity of the generalized Brink-Axel (gBA) hypothesis \cite{Brink1955, Axel1962}. 
In its most general form, the hypothesis states that the GSF is independent of excitation energies, spins and parities of initial and final states and depends solely on the $\gamma$-ray energy of involved transitions. 
This is often used as a necessary approximation and simplification in many methods, and allows to set a direct correspondence between strengths extracted from the $\gamma$-decay and photo-excitation processes \cite{Brink1955,Markova2021}. 
Even though this hypothesis is experimentally established at  high energies, \textit{i.e.}, in the vicinity of the giant dipole resonance, its validity below the neutron threshold still triggers quite some debate. 
For example, several theoretical studies suggest the need of a modification of the hypothesis \cite{BrownLarsen2014,Misch2014,Johnson2015,Hung17,Herrera22}, while experimental studies range from claims of a violation \cite{Angell12,Isaak13,Netterdon15,Isaak2019} to a validity \cite{Guttormsen2016,Martin2017,Campo2018,Scholz20,Markova2021}.
The question regarding the validity is a rather difficult one, partially depending on what degree of violation is considered acceptable in different experimental and theoretical applications.

A crucial point to be considered when addressing the applicability of the gBA hypothesis is the presence of fluctuations of partial radiative widths, or the so-called Porter-Thomas (PT) fluctuations \cite{Porter1956}. 
The partial radiative widths are proportional to the corresponding reduced transition strengths ($B(XL)$ values where $X$ is the electromagnetic character and $L$ the angular momentum of the $\gamma$ ray). 
At sufficiently high excitation energies and high NLD values, the nuclear wave functions are quite complex with many components.  
In this region, according to random-matrix theory \cite{Weidenmuller2009}, the partial widths follow a $\chi^2_{\nu}$ behavior with $\nu=1$ degree of freedom, while the total widths are more narrowly distributed with the variance inversely proportional to the number of independently contributing partial widths. 

Such a variation of partial widths is directly reflected in the variation of the GSF, which may mask the excitation energy independence of the strength, and thus a test of the gBA hypothesis might become especially difficult. 
Indeed, for relatively light nuclei, \textit{e.g.} $^{64,65}$Ni \cite{Campo2018} and $^{46}$Ti \cite{Guttormsen2011_BA}, the NLDs are rather low, and tests of the gBA hypothesis are limited. 
On the other hand, the $^{238}$Np nucleus~\cite{Guttormsen2016} with a particularly high NLD, makes a perfect case for studying the GSF as a function of initial and final excitation energies, as fluctuations of the strength are strongly suppressed. The Sn isotopes investigated here present an intermediate case for studying to what degree the PT fluctuations are expected to distort excitation energy dependence of the GSF.

Moving away from the valley of stability opens up new perspectives for studying exotic, neutron-rich nuclei, with applications to heavy-element nucleosynthesis~\cite{Larsen2019}, using for example the $\beta$-Oslo method~\cite{Spyrou2014} and the Oslo method in inverse kinematics~\cite{Ingeberg2020}.
However, this leads to additional complications, such as the lack of neutron-resonance data for  normalizing the NLD and GSF from the Oslo-method data. 
Moreover, some of the light-ion induced reactions may lead to a population of a limited spin range, which might introduce additional assumptions and uncertainties when extracting the shapes of the NLD and GSF. 

A novel technique, the Shape method~\cite{Wiedeking2020}, has recently been proposed to  mend this problem. 
Applied to the primary $\gamma$-transitions to several low-lying discrete states at consecutive excitation energy bins, it allows for an independent determination of the shape of the GSF. 
Thus, the shape of the strength and the interlinked  slope of the NLD extracted with the Oslo method can be additionally constrained by the Shape method. 
However, as the latter is using data on direct decays to low-lying discrete states only, the PT fluctuations of the involved partial widths are expected to be significantly larger than for the Oslo-method GSF. 

In this work, the potential role of PT fluctuations in establishing the validity of the gBA hypothesis as well as the application of the Shape method are addressed for $^{120}$Sn and $^{124}$Sn. 
Both the Oslo method and the Shape method have been applied to the same data sets.
Experimental NLDs have been used to estimate fluctuations of the strengths for different specific initial and final excitation energies and compared with previous Oslo-method NLDs for even-even isotopes \cite{Agvaanluvsan09, Toft2010, toft2011}. 
In Sec. \ref{sec 2: experiment} the details of the experimental procedure, the application of the Oslo method (Subsec. \ref{subsec 2.1: Oslo method}) and the Shape method (Subsec. \ref{subsec 2.1: Shape method})  are presented. 
Section \ref{sec 3: NLD} focuses on the NLDs for $^{120,124}$Sn and the comparison with other experimental and theoretical results. 
In Sec. \ref{sec 4: PT} the procedure of estimating fluctuations of the strengths is presented together with the Shape method results, and the study of fluctuations and GSFs as functions of initial and final excitation energies. 
Finally, the main conclusions are summarised in Sec. \ref{sec 6: Conclusion}.

\section{\label{sec 2: experiment}Details of the experiment and data analysis}

Experiments on both $^{120}$Sn and $^{124}$Sn were performed in February 2019 at the Oslo Cyclotron Laboratory (OCL). 
The isotopes were studied through the inelastic scattering reactions  $^{120,124}$Sn$(p, p^{\prime}\gamma$)  with a proton beam of energy 16 MeV  and intensity $I \approx 3- 4$ nA provided by the MC-35 Scanditronix cyclotron. 
Both targets used in the experiment were self-supporting with thicknesses and enrichments of 2.0 mg/cm$^2$, 99.6\% for $^{120}$Sn and 0.47 mg/cm$^2$, 95.3\% for $^{124}$Sn, respectively. 
The $^{120}$Sn target was placed in the beam for approximately 24 hours, while the whole run on $^{124}$Sn lasted approximately 17 hours. 
A self-supporting $^{28}$Si target (natural Si, 92.2\% $^{28}$Si) with thickness of $4$ mg/cm$^{2}$ was placed in the same proton beam for $\approx 1.5$ hours at the end of the experiment for the energy calibration of the $\gamma$ detectors.

The experimental setup at the OCL comprises of the target chamber surrounded by 30 cylindrical large-volume LaBr$_3$(Ce) detectors (Oslo SCintillator ARray, OSCAR for short) \cite{Ingeberg2020, Zeiser2020}, and 64 Si particle $\Delta E-E$ telescopes (SiRi) \cite{Guttormsen2011}.
The LaBr$_3$(Ce) scintillator detectors with $\diameter$3.5$^{\prime\prime}\times$8$^{\prime\prime}$ crystals were mounted on a truncated icosahedron frame with all front-ends fixed at a distance of 16.3 cm from the center of the target chamber, thus covering $\approx 57$\% of the total solid angle. 
The full-energy peak efficiency and energy resolution of OSCAR have been measured to be $\approx 20\%$ and $\approx 2.7\%$, respectively, at $E_\gamma = 662$ keV for the $^{137}$Cs calibration source placed at the same distance from the front-ends of the detectors. 

\begin{figure}[]
\includegraphics[width=1.0\columnwidth]{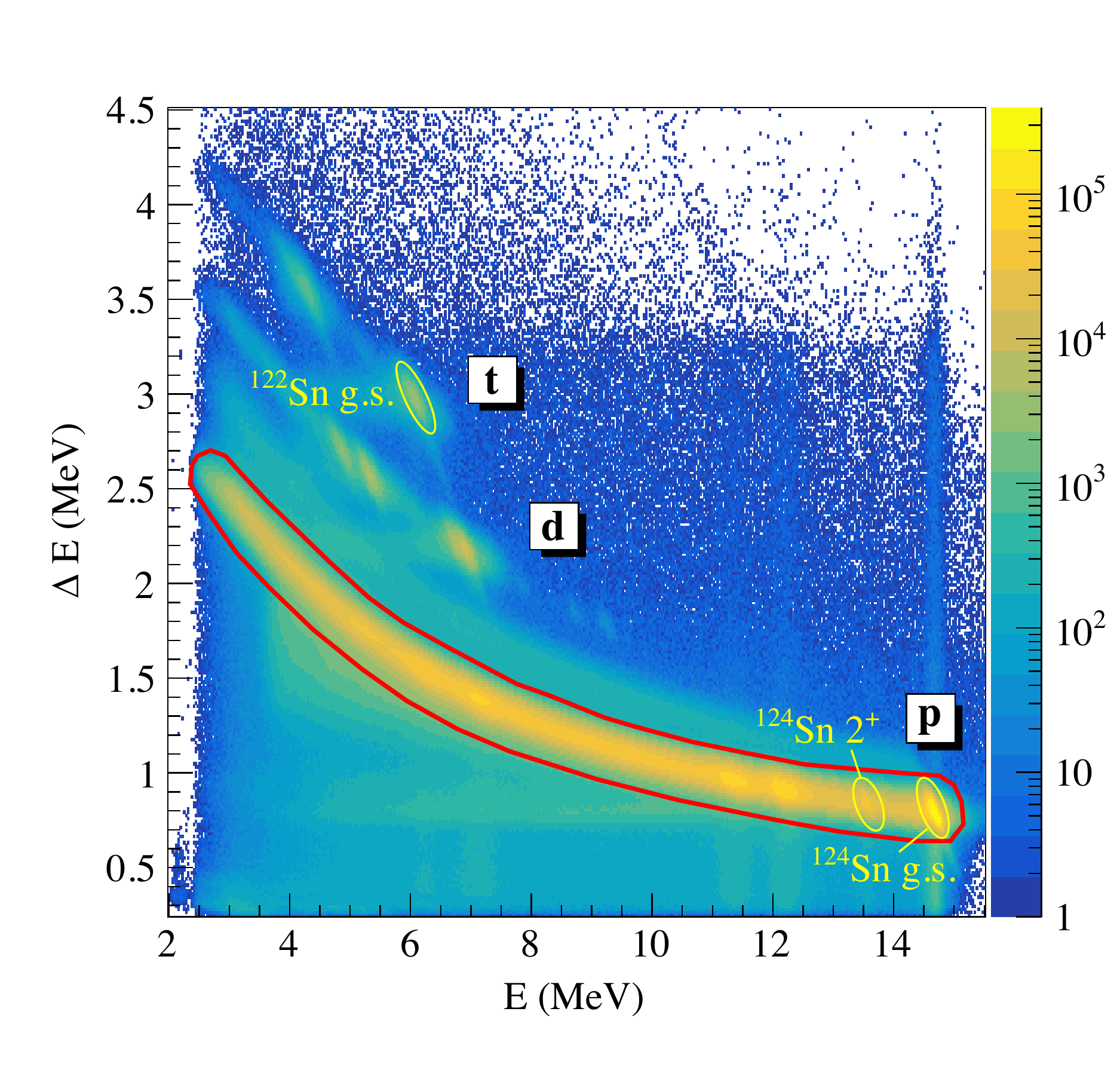}
\caption{\label{fig: E DE}
Experimental E-$\Delta$E spectrum measured for the $^{124}$Sn isotope. The proton channel used for the data analysis is marked with the red solid line. The ground and first excited states of $^{124}$Sn in the proton channel and the ground state of $^{122}$Sn in the triton channel, used for the calibration of the particle telescope, are marked with yellow circles.
}
\end{figure}

The $(p,p^{\prime}\gamma$) reaction on $^{120,124}$Sn was one of the first in the series of experiments performed with OSCAR, installed in 2018 at the OCL. 
As compared to the previously used array CACTUS, consisting of 28 5$^{\prime\prime}\times$5$^{\prime\prime}$ NaI(Tl) detectors \cite{Guttormsen1990_CACTUS}, OSCAR provides greatly improved timing and $\gamma-$energy resolution. 
All the scintillator crystals in the OSCAR array are coupled to Hamamatsu R10233-100 photomultiplier tubes with active voltage dividers (LABRVD) \cite{Riboldi2011}. 
\begin{figure*}[t]
\includegraphics[width=1.0\textwidth]{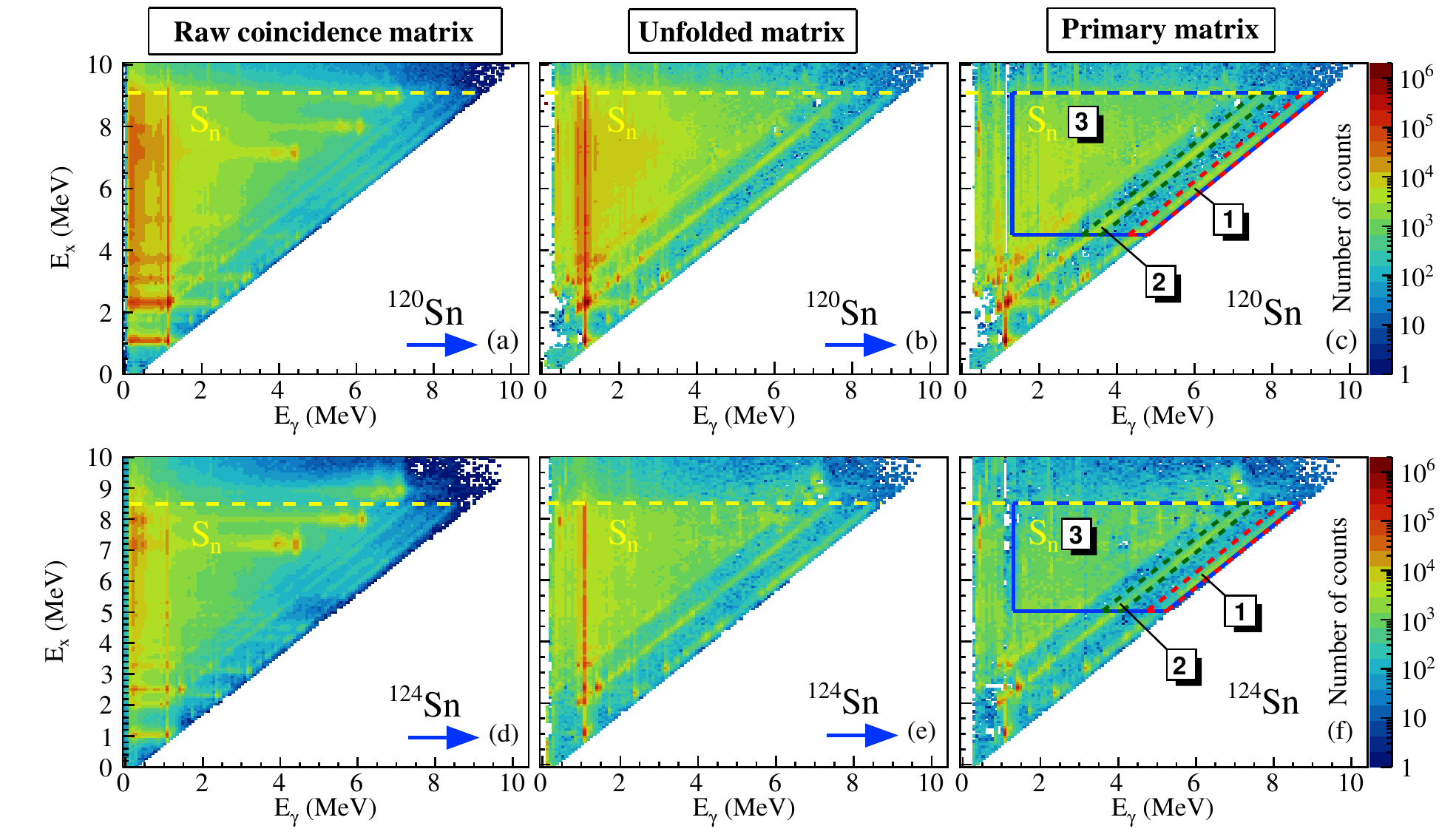}
\caption{\label{fig: matrices}  Experimental raw $p-\gamma$ coincidence ((a) and (d)), unfolded ((b) and (e)) and primary ((c) and (f)) matrices for $^{120,124}$Sn obtained in the ($p,p^{\prime}\gamma$) experiments. 
Yellow dashed lines indicate the neutron separation energies. 
Red and green dashed lines in panels (c) and (f) confine transitions to the ground (region 1) and the first excited $J^{\pi}=2^+$ (region 2) states. 
Blue solid lines (region 3) indicate the areas of the primary matrices used further in the Oslo method. 
Bin sizes are 64 keV$\times$64 keV and 80 keV$\times$80 keV for $^{120}$Sn and $^{124}$Sn, respectively. 
Blue arrows mark the sequence of the analysis steps.}
\end{figure*}

In these experiments, the SiRi particle-telescope array was placed in backward angles with respect to the beam direction, covering a rather narrow range of angles from 126$^{\circ}$ to 140$^{\circ}$ and making up $\approx6\%$ of the total solid-angle coverage. 
SiRi consists of eight trapezoidal-shaped $\Delta E-E$ telescopes with a thick $E$-detector and a thinner $\Delta E$-detector with thicknesses of 1550 $\mu$m and 130 $\mu$m, respectively. 
Each of the eight $\Delta E$ detectors is segmented into eight curved pads, amounting to 2$^{\circ}$ of particle scattering angle per pad, yielding an angular resolution of 2$^{\circ}$. 
For the $^{120,124}$Sn$(p,p^{\prime}\gamma$) experiment, the full width at half maximum (FWHM) for SiRi was estimated to be $\approx100-120$ keV from a Gaussian fit to the elastically scattered protons. 
All particle-$\gamma$ coincidences in the experiment were recorded using XIA digital electronics \cite{xia}.

SiRi enables the exploitation of the $\Delta E-E$ technique  to differentiate between the various observed reaction channels, as shown in Fig.~\ref{fig: E DE}. 
The elastic peak in the proton channel and the ground-state peak in the triton channel, combined with the known energy deposition in each of the 64 $\Delta E-E$ pads, were used to perform a linear calibration of the SiRi detectors for both targets.
The kinematics of the reactions were used to convert the proton energies deposited in the SiRi detectors into the corresponding excitation energies of the target nucleus.

As previously shown for $\diameter$3.5$^{\prime\prime}\times$8$^{\prime\prime}$ LaBr$_3$(Ce) detectors coupled to the same type of  photomultiplier and voltage divider, the energy response of the detector remains rather linear up to $\approx 17-18$ MeV \cite{Giaz2013}. 
However, to account for minor non-linearity effects, a quadratic calibration was applied to all 30 OSCAR detectors. 
Prominent $\gamma$ transitions in $^{28}$Si ranging from 1.78 to 7.93 MeV were used for this purpose.
Further, by applying graphical energy (see Fig.~\ref{fig: E DE}) and timing cuts on the studied proton channel, putting gates on the prompt timing peak and subtracting background for particle and $\gamma$ detection in SiRi and OSCAR,  a so-called raw coincidence matrix was constructed for both studied nuclei. The raw matrices are shown in Figs. \ref{fig: matrices}(a) and (d) for $^{120}$Sn and $^{124}$Sn, respectively. 
Consecutive diagonals indicate direct transitions to the ground and first excited states.
For excitation energies between 7 and 9 MeV, peaks that are due to minor $^{12}$C and $^{16}$O contaminants in the targets are observed. 
At further stages of the analysis these peaks were removed\footnote{The contaminants were removed after unfolding of the $\gamma$ spectra. A narrow graphical gate is put on each Gaussian-like contaminant peak in the unfolded matrix, and the parts of the spectra within the gate are obtained by interpolating the neighbouring regions of the matrix.} to minimize the effect of these contaminants and any related artefacts on the final results. 
Approximately $5.3 \times10^7$ and $1.3\times10^7$ $p$-$\gamma$ events in the excitation-energy range up to the neutron separation energy were collected for $^{120}$Sn and $^{124}$Sn, respectively. 

The $\gamma$ spectra for each excitation-energy bin of the coincidence matrices were further unfolded according to the procedure outlined in \cite{Guttormsen1996}, using the most recent response function of the OSCAR detectors \cite{response_function} simulated with the Geant4 simulation tool \cite{Agostinelli2003, Allison2006, Allison2016}. 
This procedure has been applied to a large number of Oslo-type data published throughout the past two decades, and has been repeatedly shown to provide valuable results. 
A great advantage of the method is the preservation of statistical fluctuations of the raw coincidence spectrum into the unfolded one by using the so-called Compton subtraction method~\cite{Guttormsen1996}. 
This technique strongly suppresses additional, artificial fluctuations. 
The unfolded matrices for $^{120}$Sn and $^{124}$Sn are shown in Figs. \ref{fig: matrices}(b) and (e).

The main objective of the analysis is to extract the statistical nuclear properties, namely the NLD and GSF, by exploiting their proportionality to the decay probability at each specific excitation energy and $\gamma$ energy. 
Information regarding this decay probability can be obtained by isolating the first $\gamma$ rays in a cascade emitted by the nucleus at a certain excitation energy, \textit{i.e.} primary $\gamma$ rays originating directly from the nucleus decaying from this excited state, or the so-called first-generation $\gamma$ rays. 
The unfolded matrix contains all possible generations of $\gamma$ rays emitted in every cascade from all excitation energies up to the neutron separation energy. 
The $\gamma$-ray spectra for each excitation-energy bin in the unfolded matrix are expected to contain the same $\gamma$ rays as in the lower-lying bins, in addition to the $\gamma$ rays originating from the excited states confined by this energy bin. This fact is the key for the iterative subtraction technique, the so-called first-generation method, applied to both unfolded matrices for $^{120}$Sn and $^{124}$Sn. This technique relies on the assumption that $\gamma$ decay is independent of whether states were populated directly in a reaction or via decays from higher-lying states. The details of the procedure are outlined in \cite{Guttormsen1987}. The primary matrices obtained after 23 iterations for both nuclei are shown in Figs. \ref{fig: matrices}(c) and (f). 

At this stage, two alternative methods can be used in order to extract the GSF from the primary matrix, namely the Oslo method and the Shape method. 
The former is a well-developed procedure primarily used to extract nuclear properties from the OCL data and it has been in use for more than two decades (see \textit{e.g.} \cite{Larsen11}). In addition to the GSF, it provides the simultaneous extraction of the NLD, which are the main characteristics of interest in this article. 
The latter procedure, the Shape method, has been recently presented and published in Ref.~\cite{Wiedeking2020}. 
The two methods are expected to complement each other and a combined analysis yields an improved normalisation of the GSF and, therefore, the NLD. 
All details of these procedures applied to the $^{120,124}$Sn isotopes are provided in the subsequent sections.


\subsection{\label{subsec 2.1: Oslo method}Analysis with the Oslo method}

As already mentioned, the primary matrix is proportional to the decay probability from a set of initial excited states $i$ within a chosen bin $E_i$ to final states $f$ confined within a bin $E_f$ of the same size with $\gamma$ rays of energy $E_{\gamma}=E_i-E_f$. The first step of the Oslo-type of analysis is the decomposition of the primary matrix into the density of final states $\rho_f$ and the $\gamma$-ray transmission coefficient $\mathcal{T}_{i\rightarrow f}$:
\begin{equation}
\label{eq:2}
    P(E_{\gamma},E_i)\propto \mathcal{T}_{i\rightarrow f}\cdot\rho_f.
\end{equation}

Here, $\mathcal{T}_{i\rightarrow f}$, the transmission coefficient, is a function of $\gamma$-ray energy depending on both the initial and final state. The thorough derivations of this decomposition using Fermi's golden rule and the Hauser–Feshbach theory of statistical reactions as starting points can be found in Refs.~\cite{MIDTBO_1} and \cite{MIDTBO_2}, correspondingly. This relation is expected to hold for relatively high excitation energies below the neutron threshold, corresponding to the compound states and their decay \cite{Larsen11}. This energy range essentially coincides with the range of applicability of the first generation method.

This form of dependence on $E_i$, $E_f$ and $E_{\gamma}$, however, does not allow a simultaneous extraction of the transmission coefficient and NLD. 
To enable such an extraction, the gBA hypothesis is adopted as one of the central assumptions in the Oslo method \cite{Brink1955, Axel1962}. 
As mentioned previously, the gBA hypothesis suggests an independence of the GSF (and, therefore, the transmission coefficient) of spins, parities and energies of initial and final states, leading to a dependence on $\gamma$-ray energy only. 
This significantly simplifies the form of the relation given in Eq.~(\ref{eq:2}): $\mathcal{T}_{i\rightarrow f}\rightarrow \mathcal{T}(E_{\gamma}$) and $\rho_f=\rho(E_f)=\rho(E_i-E_{\gamma})$. 

In earlier applications of the Oslo method, the gBA hypothesis has been found to be adequate for the relatively low-temperature regimes studied ($T\approx 0.7-1.5$ MeV)  \cite{Guttormsen2011_BA}. 
However, as the Oslo method relies on the gBA hypothesis, it cannot be used alone to test its validity. 
To investigate the validity of the hypothesis, either comparisons of independent experimental methods \cite{Markova2021} or additional tests suggested in, \textit{e.g.}, \cite{Guttormsen2016,Campo2018} are required. 
This matter is of particular importance and will be discussed in more detail in Sec.~\ref{sec 4: PT}.

The next step of the Oslo method includes an iterative $\chi^2$-minimization procedure between the experimental first-generation matrix $P(E_{\gamma}, E_i)$ normalized to unity for each $E_i$ bin and the theoretical $P_{th}(E_{\gamma}, E_i)$ given by the following expression~ \cite{Schiller2000}:
\begin{equation}
\label{eq:3}
P_{th}(E_{\gamma}, E_i)=\frac{\mathcal{T}(E_{\gamma})\rho(E_i-E_{\gamma})}{\sum_{E_{\gamma}=E_{\gamma}^{min}}^{E_i}\mathcal{T}(E_{\gamma})\rho(E_i-E_{\gamma})}.
\end{equation}
This $\chi^2$ fit of the transmission coefficient and NLD normally gives a very good agreement with the experimental matrix $P(E_{\gamma}, E_i)$ when applied to the statistical region of excitation energies. 
The step-by-step description of the minimization procedure is provided in Ref.~\cite{Schiller2000}. 
To ensure the applicability of the statistical assumptions, minimum excitation energies of $E_i^{min}=4.5$ MeV for $^{120}$Sn and $5.0$ MeV for $^{124}$Sn were chosen. 
Sufficient statistics at higher energies allows us to set $E_i^{max}$ to the neutron separation energy for each isotope, $S_n =9.1$ and 8.5 MeV for $^{120}$Sn and $^{124}$Sn, respectively. 
To exclude regions where counts have been over-subtracted in the first-generation procedure, minimum $\gamma$-ray energies $E_{\gamma}^{min}=1.3$ and 1.6 MeV were set accordingly for $^{120}$Sn and $^{124}$Sn. The resulting areas where the Oslo method was applied in this work are marked by the blue lines in Figs. \ref{fig: matrices}(c) and (f).

The global $\chi^2$ fit yields only functional forms of the transmission coefficient $\mathcal{T}(E_{\gamma})$ and NLD $\rho(E_i-E_{\gamma})$. 
It can be shown mathematically that one can construct an infinite set of $\mathcal{T}(E_{\gamma})$ and $\rho(E_i-E_{\gamma})$ combinations corresponding to the obtained fit and given by the forms \cite{Schiller2000}: 
\begin{equation}
\label{eq:4}
\begin{split}
    \tilde{\rho}(E_i-E_{\gamma})=&A\rho(E_i-E_{\gamma})\exp(\alpha (E_i-E_{\gamma})),\\
    \tilde{\mathcal{T}}(E_\gamma)=&B\mathcal{T}(E_\gamma)\exp(\alpha E_{\gamma}),
\end{split}
\end{equation}
where $\rho$ and $\mathcal{T}$ are two fixed solutions, $A$ and $B$ are the scaling parameters, and $\alpha$ is the slope parameter shared by both the transmission coefficient and NLD. 
For each studied nucleus this ambiguity must be removed via determining unique normalization parameters $A,B$ and $\alpha$ from  external experimental data. 
If available, low-lying discrete states and neutron-resonance data are the main input parameters, combined with  models for the spin distribution and for extrapolations where there is a lack of experimental data. 

The first step of the normalization procedure is to determine the unique NLD solution $\rho(E_i-E_{\gamma})$. 
The parameters $A$ and $\alpha$ can be constrained by fitting the NLD to low-lying discrete states \cite{ensdf} in the excitation-energy range where the level scheme can be considered complete. 
At the neutron separation energy, the NLD can be normalized to the total NLD calculated from  neutron-resonance spacings \cite{Mughabghab18}. 
These data also provide the average, total radiative width $\langle \Gamma_{\gamma}\rangle$ used to determine the scaling parameter $B$ for the transmission coefficient. 
All details of the normalization procedure for $^{120}$Sn and $^{124}$Sn have been presented in the Supplemental Material of Ref.~\cite{Markova2021}. 
However, some minor changes were introduced in this work to improve the normalization and the estimated uncertainties. 
We would like to stress that these changes do not affect the results presented in Ref.~\cite{Markova2021} in any significant way, and do not undermine any of the presented conclusions. 
To avoid any confusion regarding the normalization parameters, we provide the updated and complete description of this procedure in the following.

\begin{figure}[]
\includegraphics[width=1.0\columnwidth]{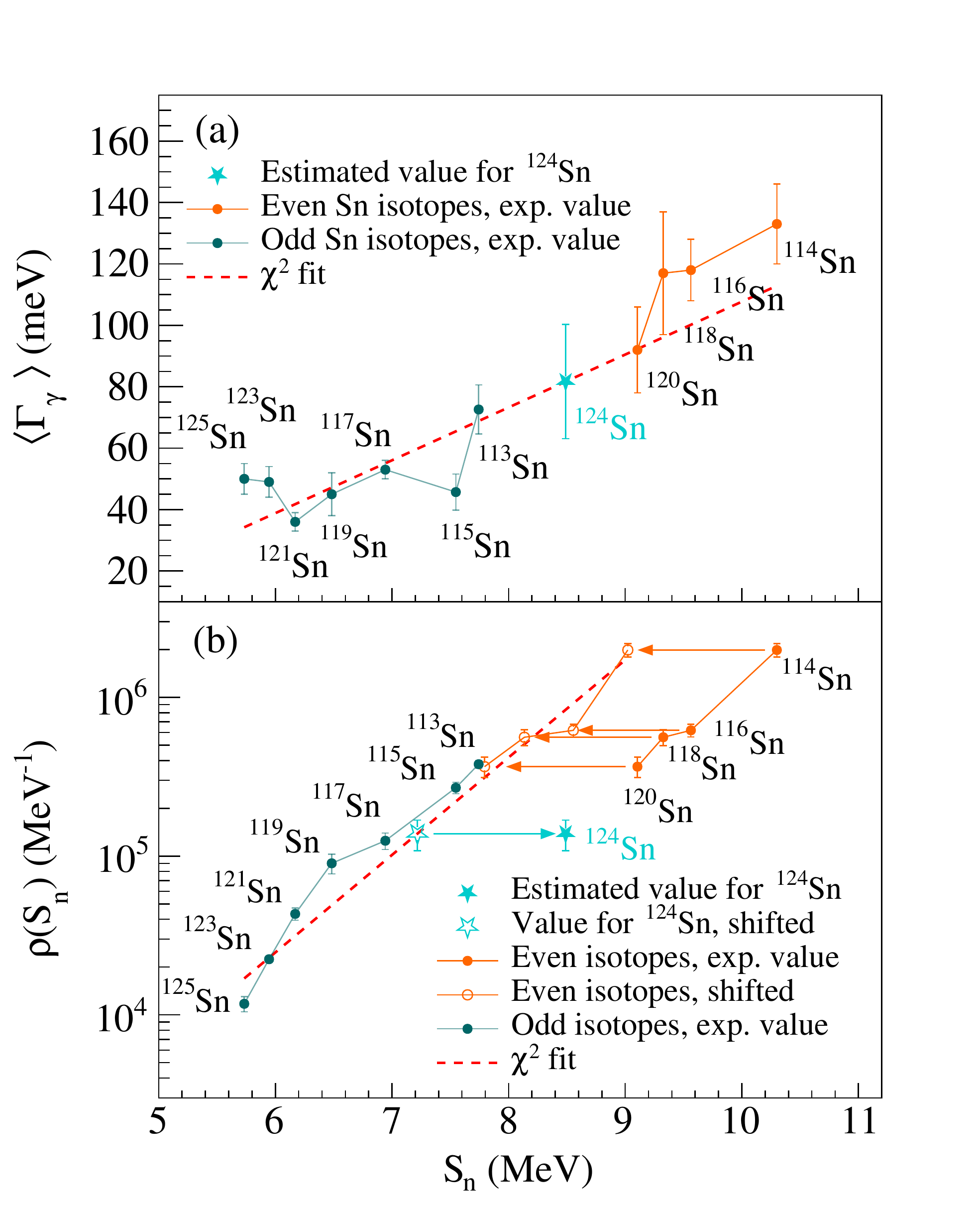}
\caption{\label{fig: systematics}
(a) Experimental systematics for the average total radiative width for Sn isotopes.
(b) Experimental systematics for the NLD at the neutron separation energy. 
The estimated values of $\langle \Gamma_{\gamma}\rangle$ and $\rho(S_n)$ for $^{124}$Sn are marked with stars, the experimental $\langle \Gamma_{\gamma}\rangle$ values are taken from \cite{Mughabghab18}, the level densities are obtained from the $D_0$ values given in \cite{Mughabghab18}. Arrows mark $\rho(S_n)$ values shifted by the neutron pair-gap values for the $\chi^2$ fit.
}
\end{figure}

The most recent compilation of the discrete states \cite{ensdf} was used to anchor the NLD for $^{120,124}$Sn at low excitation energies. 
As compared to the compilation from 2003 used in the previous analysis, some changes in the number and the excitation energies of low-lying states appear and give a slightly different slope of the NLD. 
The anchor point at the neutron separation energy, $\rho(S_n)$, is usually extracted from the neutron resonance spacing $D_0$ for $s$-wave neutrons or $D_1$ for $p$-wave neutrons. 
As $^{123}$Sn is an unstable target nucleus ($T_{1/2}=129.2$ d \cite{ensdf}), no neutron resonance data are available, and we used other means to estimate $\rho(S_n)$ and $\langle\Gamma_{\gamma}\rangle$ for $^{124}$Sn.

The normalization procedure for $^{120}$Sn is rather straightforward, in accordance with the steps outlined in Ref.~\cite{Larsen11}, due to the available $s$-wave neutron capture data. 
The target spin of $^{119}$Sn is $I_{t}^{\pi}=1/2^+$, thus spins $0^+$ and $1^+$ of the compound nucleus $^{120}$Sn are populated in $s$-wave capture. 
Assuming that both positive and negative parities contribute equally to $\rho(S_n)$, the average $s$-wave neutron resonance spacing $D_0$ can be written as \cite{Larsen11}:
\begin{equation}
\label{eq:5}
    \frac{1}{D_0}=\frac{1}{2}\left[\rho(S_n, I_t+1/2)+\rho(S_n, I_t-1/2)\right].
\end{equation}

A transformation of the \textit{partial} NLD for specific spins into the \textit{total} NLD can be performed by adopting the back-shifted Fermi gas model (BSFG) for the NLD $\rho(E_x,J)=\rho(E_x)\cdot g(E_x,J)$ ($E_x$ here stands for the excitation energy variable) with the spin distribution function given by \cite{Ericson58,Gilbert65}:
\begin{equation}
\label{eq:6}
 g(E_x, J) \simeq \frac{2J+1}{2\sigma^2}\exp\left[-\frac{(J+1/2)^2}{2\sigma^2}\right],
\end{equation}
where $\sigma$ is the spin-cutoff parameter. 
Given this distribution function, Eq.~(\ref{eq:6}) can be rewritten for the total NLD at the neutron separation energy as a function of the experimental resonance spacing $D_0$ (taken from Ref. \cite{Mughabghab18}) and the target nucleus spin \cite{Larsen11}:
\begin{equation}
\label{eq:7}
\rho(S_n)=\frac{2\sigma^2}{D_0}\frac{1}{(I_t+1)\exp(-\frac{(I_t+1)^2}{2\sigma^2})+I_t\exp(-\frac{I_t^2}{2\sigma^2})}.
\end{equation}
Note that the spin-cutoff parameter is an excitation-energy dependent function. 
The form of the spin-cutoff parameter at $S_n$ of Ref. \cite{Egidy05} was chosen for $^{120,124}$Sn
\begin{equation}
\label{eq:8}
\sigma^2(S_n)=0.0146A^{5/3}\frac{1+\sqrt{1+4a(S_n-E_1)}}{2a}.
\end{equation}
Here, $a$ and $E_1$ are the level-density parameter and the back-shift parameter for the BSFG model taken from Ref.~\cite{Egidy05}.

In the Oslo method, the measured level densities do not reach up to  $E_x=S_n$  due to the non-zero minimum $\gamma$-ray energy limit in the extraction of $\rho(E_i-E_\gamma)$. 
To use the $\rho(S_n)$ value as an anchor point for the normalization, the experimental Oslo data were extrapolated using the constant temperature (CT) level density model \cite{Ericson59,Gilbert65,Egidy05}:
\begin{equation}
\label{eq:9}
    \rho_{CT}(E_x)=\frac{1}{T_{{CT}}}\exp(\frac{E_x-E_0}{T_{{CT}}}),
\end{equation}
characterized by temperature ($T_{CT}$) and shift energy ($E_0$) parameters. 
Earlier Oslo-method analyses exploited the BSFG model as an alternative for the interpolation procedure \cite{Larsen11}, however, the choice between these two alternatives is defined by the fit quality in each particular case (see Sec.\ref{sec 3: NLD}).

As the experimental information on the $s$-wave neutron-resonance spacing is available for $^{120}$Sn, Eq.~(\ref{eq:7}) was used directly to transform the $D_0$ value into $\rho(S_n)$. 
For $^{124}$Sn, this value was estimated from the systematics for even-even  and  even-odd Sn  isotopes  in  the  following  way. 
The $\rho(S_n)$ values were estimated for each Sn isotope with available neutron-resonance spacing $D_0$ using Eq.~(\ref{eq:7}). 
The resulting systematics for the $\rho(S_n)$ values are shown in the lower panel of Fig.~\ref{fig: systematics}. 
The values of $\rho(S_n)$ for even-even isotopes were shifted by the corresponding values of the neutron pairing gaps calculated from the AME 2003 mass evaluation \cite{ame03} using Eq.~(1) of Ref.~\cite{Dobaczewski01}. 
Finally, the value of $\rho(S_n)$ for $^{124}$Sn was calculated from a log-linear fit through the data points for even-odd and shifted even-even isotopes as shown by the red dashed line in  Fig.~\ref{fig: systematics}(b).

The second step after constraining the $A$ and $\alpha$ parameters for the NLD is to normalize the transmission coefficient (and thus the GSF). 
As the slope $\alpha$ is already determined by the NLD normalization, the scaling parameter $B$ is the only parameter that remains to be estimated. 
The starting point for normalizing the $\gamma$-transmission coefficient is the following relation~\cite{Kopecky1990}:
\begin{equation}
\label{eq:10}
\begin{split}
    \langle\Gamma(E_x, J, \pi)\rangle=&\frac{1}{2\pi\rho(E_x, J, \pi)}\sum_{XL}\sum_{J_f,\pi_f}\int_{E_{\gamma}=0}^{E_x}dE_{\gamma}\times\\&\times\mathcal{T}_{XL}(E_\gamma)\rho(E_x-E_{\gamma},J,\pi),
\end{split}
\end{equation}
where $\langle\Gamma(E_x, J, \pi)\rangle$ is the average radiative width for states with spin $J$, parity $\pi$ at  excitation energy $E_x$, and $X$ and $L$ indicate the electromagnetic character and multipolarity, respectively. 
The  GSF, $f_{XL}(E_{\gamma})$, is connected to the transmission coefficient by~\cite{Belgya2006}:
\begin{equation}
\label{eq:11}
    \mathcal{T}_{XL}(E_\gamma)=2\pi E_{\gamma}^{(2L+1)}f_{XL}(E_{\gamma}).
\end{equation}

At high excitation energies, there is experimental evidence  the dipole radiation is dominant $(L=1)$ (\textit{e.g.}, Ref.~\cite{Kopecky1990}). The Oslo-type of experiments and analysis does not allow for distinguishing between different types of radiation, and, thus, the strength extracted with the Oslo method is presented by the total contribution of both electric and magnetic types of dipole transitions, $E1$ and $M1$.

Insertion into Eq.~(\ref{eq:10}) links the experimental dipole GSF $f(E_{\gamma})$ to the value of the total average radiative  width $\langle\Gamma_{\gamma}\rangle$ obtained from $s$-wave neutron capture \cite{Mughabghab18}. 
For a target nucleus with ground state spin $I_t$ and parity $\pi_t$, Eq.~(\ref{eq:10})  can be rewritten as
\begin{equation}
\label{eq:12}   
\begin{split}
    &\langle\Gamma_{\gamma}\rangle=\langle\Gamma(S_n, I_t\pm1/2, \pi_t)\rangle=\frac{1}{2\rho(S_n, I_t\pm1/2, \pi_t)}\times\\&\times\int_{E_{\gamma}=0}^{S_n}dE_{\gamma} E_{\gamma}^3f(E_\gamma)\rho(S_n-E_{\gamma})\times\\&\times\sum_{J=-1}^{1}g(S_n-E_{\gamma},I_t\pm1/2+J).
\end{split}
\end{equation}
Here, we adopt again the assumption on an equal contribution of states with positive and negative parities, and apply the spin distribution function of Eq. (\ref{eq:6}). 
It can be easily seen that the $1/\rho(S_n, I_t\pm1/2,\pi_t)$ term equals the $D_0$ value. 
For the spin-cutoff parameter dependence on the excitation energy, we follow the procedure outlined in Ref.~\cite{Capote2009}:
\begin{equation}
\label{eq:13}  
    \sigma^2(E_x) = \sigma_d^2 + \frac{E_x-E_d}{S_n-E_d}[\sigma^2(S_n)-\sigma_d^2],
\end{equation}
where $\sigma_d$ is estimated from the discrete lower-lying states at $E_x\approx E_d$~\cite{ensdf} (see Table~\ref{tab:table_1}).

In the case of $^{120}$Sn, the average total radiative width $\langle\Gamma_{\gamma}\rangle$ was estimated as an average of three $s$-wave  neutron  resonances with energies in the range of $\approx 455$-828~eV~\cite{Mughabghab18}.
The remaining two resonances presented in \cite{Mughabghab18} were excluded due to either being possibly of $p$-wave nature, or having a significantly lower value as compared to values for confirmed $s$-wave resonances found in the neighbouring Sn isotopes. 
In the case of $^{124}$Sn, we performed a linear fit through all values of $\left< \Gamma_{\gamma}\right>$ available for other Sn isotopes as shown in Fig.~\ref{fig: systematics}(a) to estimate the $\left<\Gamma_{\gamma}\right>$ value for $^{124}$Sn.

\begin{figure}[]
\includegraphics[width=1.0\columnwidth]{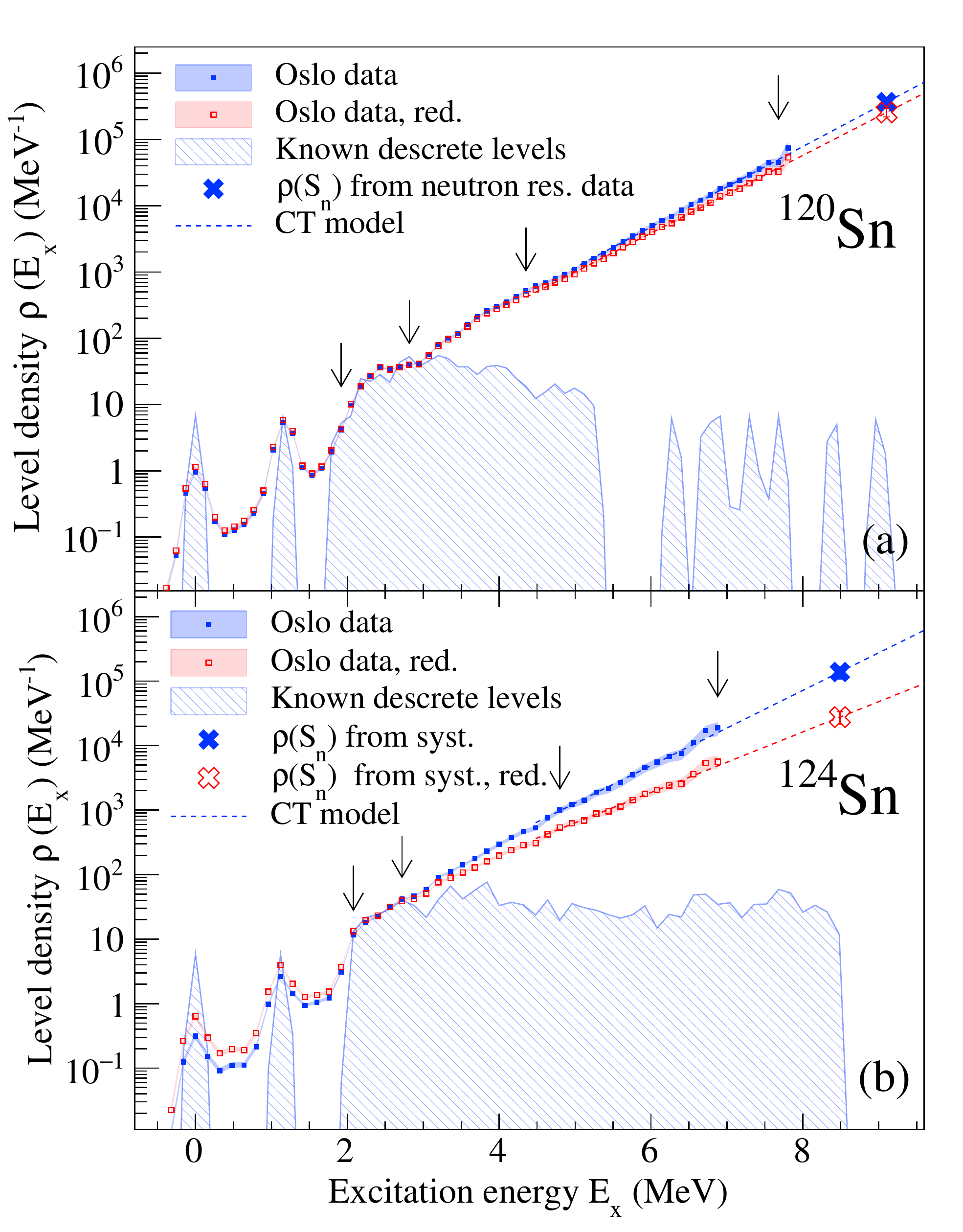}
\caption{\label{fig: LD}
Experimental nuclear level densities for $^{120}$Sn (a) and $^{124}$Sn (b). The NLDs at $S_n$ are marked with crosses, discrete states are shown as shaded areas. For the $^{124}$Sn isotope both the total and reduced NLDs are shown. The first two vertical arrows at lower $E_x$ energies on each figure constrain the lower excitation energy fit region, while the last two arrows at higher $E_x$ energies mark the lower and upper limits for the higher excitation energy fit region.
}
\end{figure}

\begin{table*}[t]
\caption{\label{tab:table_1}Parameters used for the normalization of the nuclear LD and GSF for $^{120,124}$Sn.}
\begin{ruledtabular}
\begin{tabular}{lccccccccccccc}
Nucleus & $S_n$ & $D_0$ & $a$ & $E_1$ & $E_d$ & $\sigma_d$ & $\sigma(S_n)$ & $\rho(S_n)$ & $T$ & $E_0$ & $\beta$ & $\langle\Gamma_{\gamma}\rangle$ \\ 
& (MeV) & (eV) & (MeV$^{-1}$) & (MeV) & (MeV) & & & ($ 10^5$ MeV $^{-1}$) &  (MeV) & (MeV) & & (meV)\\
\noalign{\smallskip}\hline\noalign{\smallskip}
 $^{120}$Sn & 9.105 & 95(14) & 13.92 & 1.12 & 2.53(4) & 3.4(5) & 5.82 & 3.66(54) & 0.72$^{+1}_{-2}$ & 0.19$^{+9}_{-4}$ & 0.70 &  121(25)\footnotemark[2] \\
 $^{124}$Sn & 8.489 & -- & 12.92 & 1.03 & 2.77(3) & 3.3(5) & 6.00 & 1.38(30)\footnotemark[1] & 0.75$^{+2}_{-2}$ & -0.11$^{+11}_{-6}$ & 0.20 & 82(19)\footnotemark[1] \\
\end{tabular}
\end{ruledtabular}
\footnotetext[1]{From systematics.}
\footnotetext[2]{Modified with respect to the value pubished in Ref. \cite{Mughabghab18}.}
\end{table*}

Ideally, the fit of the NLD to the low-lying discrete levels and the $\rho(S_n)$ value are sufficient to constrain the slope parameter $\alpha$ for the GSF and NLD. 
However, the latter can be influenced by the range of experimentally populated spins, which might be narrower than the \textit{intrinsic\footnote{All existing spins possible for a given nucleus at a given excitation energy.}} spin distribution. 
This issue was previously discussed in Refs.~\cite{Zeiser2018a,Zeiser2019}. 
An analysis of the observed transitions in the unfolded matrices below $E_i\approx 4-5$ MeV and their relative intensities can aid to reveal the populated spins of the $^{120,124}$Sn nuclei populated in the $(p,p^{\prime}\gamma)$ reaction. 
However, this method has a large uncertainty in the determination of the exact maximum spin populated in the reaction. 
Alternatively, one can make use of the new Shape method~\cite{Wiedeking2020} to obtain the NLD slope that corresponds to the experimental spin range. 
This is of particular importance for $^{124}$Sn with no available neutron-resonance parameters. 
The application of the Shape method will be discussed in detail in Sec.~\ref{subsec 2.1: Shape method}.
From the Shape method we obtained a reduction factor $\beta$ for $\rho(S_n)$, representing a certain fraction of the total spin distribution from Eq.~(\ref{eq:6}), corresponding to the reduced spin range from $J=0$ to a certain maximum spin. This was done by requesting optimally matching slopes of the Oslo method and the Shape method GSFs above $E_{\gamma}\approx 5$ MeV. 
A rather strong reduction of the level density in $^{124}$Sn at the neutron separation energy might reflect some maximum limit of the experimental spin range. 
However, it is important to note that using  experimental systematics of the $\rho(S_n)$ and $\langle\Gamma_{\gamma}\rangle$ might have large uncertainties. 
In the case of $^{124}$Sn, it is quite probable that such a large reduction factor is needed due to, \textit{e.g.}, an overestimated $\rho(S_n)$ from the $\chi^2$ fit of the systematics. 
The simultaneous use of the Oslo and Shape methods can therefore significantly reduce systematic uncertainties for the slopes of extracted strengths and level densities. 
All parameters used in the normalization procedure for $^{120,124}$Sn are listed in Table \ref{tab:table_1}. 
The resulting NLDs for $^{120}$Sn and $^{124}$Sn with their estimated error bands are shown in Fig.~\ref{fig: LD}.

We note that the errors in Table~\ref{tab:table_1} and the resulting error bands for the NLD and the GSF presented in sections \ref{sec 3: NLD} and \ref{sec 4: PT} combine  statistical and systematic components. 
The latter includes uncertainties introduced by the unfolding and the first-generation procedures for both $^{120,124}$Sn isotopes. 
These are propagated through the Oslo method according to the procedure outlined in Ref.~\cite{Schiller2000}. 
In addition, systematic uncertainties due to the normalization parameters are included. 
For the $^{120}$Sn isotope, the experimental uncertainty (1 standard deviation) of the $D_0$ value was propagated to estimate the error for the NLD at the neutron separation energy. 
The experimental uncertainties of the radiative widths in $^{120}$Sn \cite{Mughabghab18} were used to estimate the error of the average, total radiative width $\langle\Gamma_{\gamma}\rangle$, contributing to the uncertainty of the scaling factor $B$. 
In the case of the $^{124}$Sn isotope, the errors of the $\rho(S_n)$ and $\langle\Gamma_{\gamma}\rangle$ were calculated from the uncertainties of the $\chi^2$ fit parameters and propagated  into the total uncertainties of the NLD and GSF. 
In the previously published result on $^{124}$Sn \cite{Markova2021}, a 50\% uncertainty of $\rho(S_n)$ was assumed to account for presumably underestimated errors from the $\chi^2$ fit. 
However, the excellent agreement within the estimated error bands of the slopes of the GSFs obtained with the Oslo and Shape method allows us to apply a more modest error band as presented in this work. 
All errors of the normalization parameters described above are summarised in Table \ref{tab:table_1}.

\subsection{\label{subsec 2.1: Shape method}Analysis with the Shape method}

Quite often, nuclei with no available neutron resonance data  and/or a restricted experimental spin range are encountered. 
One possible way to overcome this is the use of isotopic systematics comprising of nuclei with stable neighbouring $A-1$ isotopes as applied in the present case for $^{124}$Sn.
However, this is often not possible in other isotopic chains due to the lack of data (\textit{e.g.} $^{127}$Sb \cite{Pogliano2022}). 
Moreover, the question on whether systematics from neighbouring isotopic chains can be used for a given nucleus, and to what extent one can rely on these systematics, is still open. 
Hence, an alternative way to constrain the normalization parameters is required. 
The novel Shape method~\cite{Wiedeking2020} provides a way to determine the slope parameter $\alpha$ for the NLD and the GSF without making use of neutron resonance data.

The starting point for the method is extracting experimental intensities of first generation $\gamma$ transitions to specific final states with spins and parities $J^{\pi}$ at final excitation energies $E_f$, represented by diagonals in the primary matrix. 
The intensities (related to the branching ratios) of these $\gamma$ transitions are proportional to the number of counts $N_D$ in the diagonals. 
The selection of which diagonals are to be used depends on a particular nucleus, the spacing between the final states, and whether the resolution is sufficient to distinguish between different diagonals. 
The main concept behind the Shape method is that the intensities of the $\gamma$ transitions are proportional to the partial widths and hence to the GSF. 
By taking intensities of transitions in successive excitation energy bins, the functional form of the GSF can be obtained.

In the case of  $^{120,124}$Sn, the only two diagonals clearly seen in the primary matrices are the ground state diagonal $D_1$ and the diagonal corresponding to the first excited state $D_2$ (marked accordingly as regions 1 and 2 in Fig.~\ref{fig: matrices}(c) and (f)). 
For given initial excitation-energy bins $E_i$ (horizontal line) they define the direct decay to the final excitation energy $E_f$ at the ground state with $J^{\pi}=0^+$ and the first excited state with $J^{\pi}=2^+$ with  $\gamma$-ray energies $E_{\gamma}=E_i-E_f$.

The Shape method adopts the same form of the spin distribution, given by Eq.~(\ref{eq:6}), as used in the Oslo method, and assumes $\gamma$ transitions to be of predominantly dipole nature (this has been confirmed by measuring angular distributions). 
According to  Eq.~(13) in Ref.~\cite{Wiedeking2020}, the number of counts in a chosen diagonal $N_D$ corresponding to the final energy $E_f$ is proportional to the population cross-section of initial states $E_i$ with $J_i=J_f-1, J_f, J_f+1$, spin distribution function $g(E_i, J_i)$ and the partial $\gamma$-decay width. 
For the case of $^{120,124}$Sn with the ground and first excited state diagonals $D_1$ and $D_2$, the following relations can be written
\begin{equation}
\label{eq:14} 
\begin{split}
    f(E_{\gamma 1})&\propto\frac{N_{D_1}}{E_{\gamma 1}^3\cdot g(E_i, 1)}\\
f(E_{\gamma 2})&\propto\frac{N_{D_2}}{E_{\gamma 2}^3\cdot[g(E_i, 1)+g(E_i, 2)+ g(E_i, 3)]}.
\end{split}
\end{equation}
By varying $E_i$, one obtains corresponding pairs of values $f(E_{\gamma}=E_i)$ and $f(E_{\gamma}=E_i-E_x(2^+))$. As  Eqs.~(\ref{eq:14}) only give the proportionality with the GSF, these pairs are not normalized in absolute value. 

Firstly, the consecutive pairs of values are normalized internally, as shown and described in Fig. 2 of Ref.~\cite{Wiedeking2020}, to reconstruct the functional shape of the GSF. 
Thus, one can extract two GSFs, corresponding to decays to the ground state and decays to the first excited state. 
Secondly, the general shape of both GSF must be scaled to match any available strength below the neutron separation energy, \textit{i.e.} normalizing to  external experimental data. 
This is  the main limitation of the method, as it provides only a  slope, or a shape of the strength, but not the absolute GSF, and therefore requires some additional experimental information. 
For the $^{120,124}$Sn isotopes, the GSFs extracted from relativistic Coulomb excitation in forward-angle inelastic
proton scattering below the neutron separation energy \cite{Bassauer2020b} were used to scale the GSF points obtained for both diagonals separately \cite{Markova2021}. 

The upper excitation energy limit for the application of the Shape method 
can, in principle, be extended to $S_n$, whilst the definition of the lower limit is rather arbitrary. 
The applicability of Eqs.~(\ref{eq:14}) is restricted to the statistical excitation energy region where the spin distribution function $g(E_x, J)$ can be trusted. 
There is no clear criterion for the minimum level density which can be considered high enough to assume this is fulfilled. 
In this work, we 
require that the level density must be at least 10 levels per excitation energy bin for the spin distribution $g(E_x, J)$ to be applied. 


\section{\label{sec 3: NLD}Nuclear level densities}

The experimental NLDs of $^{120, 124}$Sn displayed in Fig.~\ref{fig: LD} follow nicely the discrete low-lying states up to $\approx 3$ MeV for $^{120}$Sn and $\approx 2.7$ MeV for $^{124}$Sn. 
At higher energies, the NLDs increase rapidly and reach an exponential, constant-temperature behavior. 
This suggests that the level schemes used for the normalization of the NLDs can be considered complete up to $\approx 3$ and 2.7 MeV for $^{120}$Sn and $^{124}$Sn, respectively. 
The energy resolution is sufficient to distinguish the ground state and the first excited states, presented by two bumps at 0 and $\approx1.1-1.2$ MeV for both nuclei. The presence of the data points between the ground and first excited states can be explained by the finite excitation energy resolution of order 100 keV and presence of the leftover counts between the diagonals in the primary matrices after the background subtraction procedure.
At higher excitation energies, the experimental points are following the CT model prediction, starting from $\approx 4$ MeV. 
The normalization of the NLDs was found to be rather insensitive to the exact choice of the two upper normalization limits (the two arrows at higher excitation energies in Fig.~\ref{fig: LD}), due to the  smooth behaviour of the NLDs at higher excitation energies. 

\begin{figure}[t]
\includegraphics[width=1.0\columnwidth]{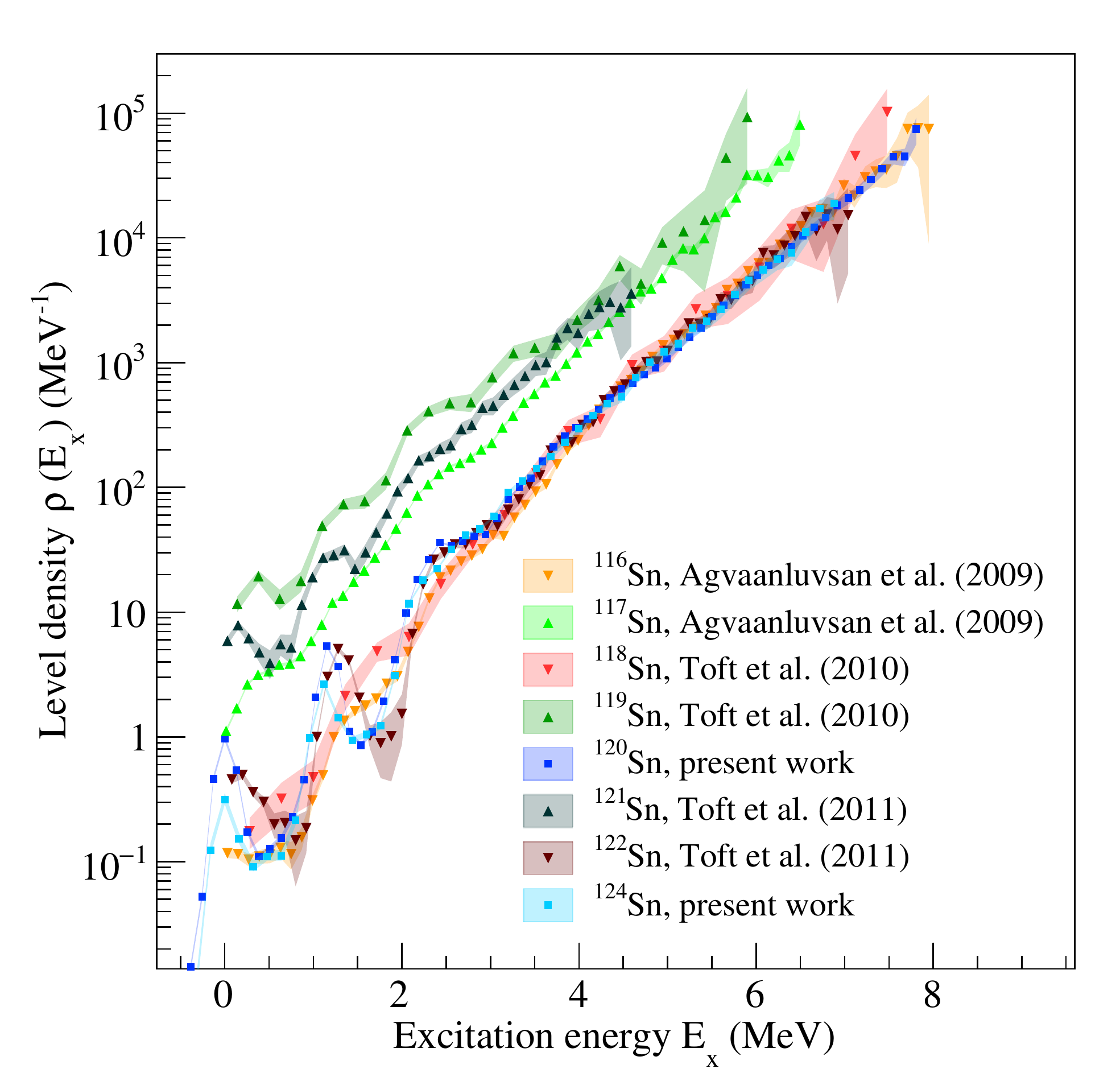}
\caption{\label{fig: LD_Sn}
Experimental total nuclear level densities for $^{116}$Sn \cite{Agvaanluvsan09}, $^{117}$Sn \cite{Agvaanluvsan09}, $^{118}$Sn \cite{Toft2010}, $^{118}$Sn \cite{Toft2010}, $^{120}$Sn, $^{121}$Sn \cite{toft2011}, $^{122}$Sn \cite{toft2011}, $^{124}$Sn.
}
\end{figure}

In Fig.~\ref{fig: LD_Sn} we show a comparison of the total NLDs for Sn isotopes extracted with the Oslo method, including the present results for $^{120,124}$Sn.
The  $^{116,117, 118, 119, 121, 122}$Sn isotopes were previously studied with a 38-MeV beam of $^3$He using the ($^3$He, $\alpha \gamma$) and ($^3$He, $^3$He $\gamma$) reaction channels and reported in Refs.~\cite{Agvaanluvsan09, Toft2010, toft2011}.
The slopes of the NLDs for $^{120,124}$Sn are quite similar to each other ($T=0.72$ and 0.75 MeV, see Table~\ref{tab:table_1}) and those of other even-mass isotopes. All NLDs of even-mass nuclei agree quite well within the estimated error bands below the neutron separation energy. However, it is important to note some differences in the normalization procedures in the newest analysis of $^{120,124}$Sn and the older analyses of even-mass isotopes.  Firstly, all previous analyses exploited the BSFG for the extrapolation of the highest experimental NLD points to the $\rho(S_n)$ values instead of the CT model. As was previously shown in Ref.~\cite{Guttormsen2015} and confirmed for $^{120,124}$Sn,  the CT model results in a better $\chi^2$ fit value. 
For example, between $\approx 4.8$ and 6.8 MeV in $^{124}$Sn, the reduced $\chi^2$ value is  a factor of 4 smaller for the CT model than for the BSFG model.
This factor becomes larger and might exceed 10 if lower excitation energy points above $\approx 3$ MeV are included. Secondly, the different form of the spin-cutoff parameter taken from Ref.~\cite{Gilbert65} was used in the older analyses. The immediate consequence of this choice is slightly less steep slopes of the NLDs if the CT extrapolation is used. However, in combination with the BSFG extrapolation model, the resulting slopes of the NLDs in $^{116,118,122}$Sn are expected to be close to those obtained for $^{120,124}$Sn, as can also be observed in Fig.~\ref{fig: LD_Sn}.

\begin{figure}[t]
\includegraphics[width=1.0\columnwidth]{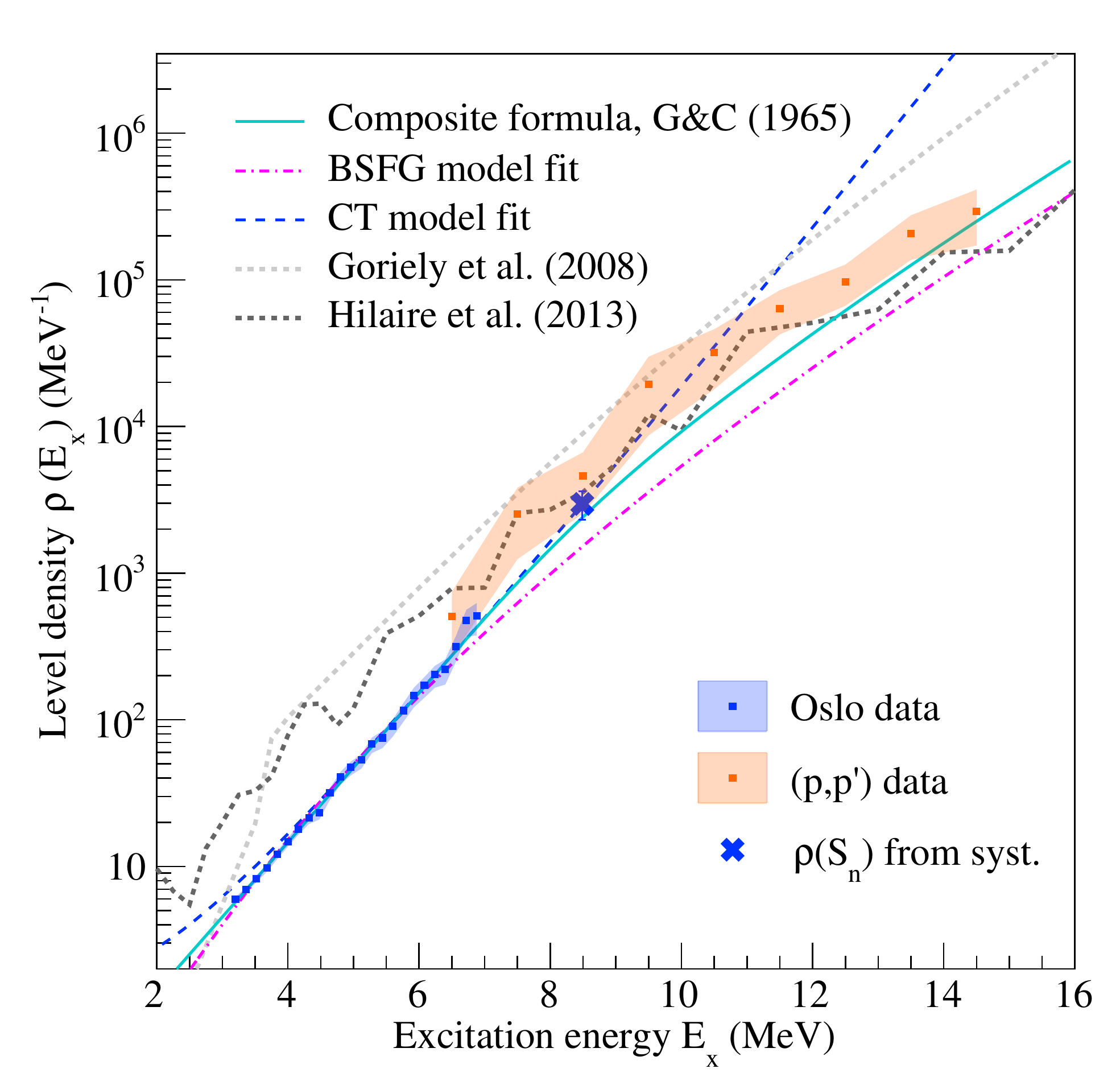}
\caption{\label{fig: LD_Oslo_Darmstadt}
Experimental nuclear level densities for 1$^{\pm}$ states  for $^{124}$Sn obtained with the Oslo method (blue data points) and the ($p,p^{\prime})$ data \cite{Bassauer2020b} (orange data points). The prediction of the CT model used for the normalization of the Oslo method data is shown by the dashed blue line. A fit with the BSFG through all data and with the composite formula \cite{Gilbert65} are shown by the dashed magenta and solid cyan lines. Predictions of the microscopic Hartree-Fock-Bogoliubov+combinatorial method \cite{Goriely2008} and Hartree-Fock-Bogolyubov+Gogny force calculations \cite{Hilaire2013} are marked by the dashed light and dark grey lines respectively.
}
\end{figure}

In general, the NLDs of odd-mass Sn isotopes are by a factor of 7-8 higher than for the even-mass isotopes, primarily due to the unpaired valence neutron~\cite{Guttormsen2003}. 
As compared to other even-mass isotopes, $^{120,124}$Sn demonstrate essentially the same features, such as the well-defined bumps at the ground and the first excited state and a step-like structure right below 3 MeV excitation energy. 
Earlier studies exploiting microscopic calculations based on the seniority model link the latter feature to breaking of consecutive nucleon Cooper pairs~\cite{Schiller2003}. 
Due to the closed proton shell, $Z=50$, the breaking of proton Cooper pairs is suppressed until higher excitation energies are reached.
Thus, these step-like structures are likely to be correlated with the breaking of neutron pairs at energies exceeding $2\Delta_n=2.6$ and 2.5 MeV \cite{Dobaczewski01} for $^{120}$Sn and $^{124}$Sn, respectively. 
For higher excitation energies, where a continuous ``melting'' of Cooper pairs sets in, the NLDs follow a smooth trend with no distinctive structures, as previously observed for $^{116, 118, 122}$Sn~\cite{Agvaanluvsan09, Toft2010, toft2011}. 

The inelastic proton scattering data \cite{Bassauer2020b}, used for the absolute normalization of the Shape method GSFs, can also provide information on the partial NLD. 
The NLD of $1^{-}$ states in $^{124}$Sn was extracted for the excitation-energy range  $\approx4.5-14.5$ MeV by means of the fluctuation analysis \cite{Hansen1990}, applying procedures analog to those used in Refs.~\cite{Poltoratska14,Bassauer2016}. All details of the extraction procedure can be found in Ref.~\cite{BassauerThesis}.
To compare with the Oslo data, we apply the spin distribution function in Eq.~(\ref{eq:6}) to the total NLD of $^{124}$Sn to reduce it to the density of $J=1$ levels for excitation energies above $\approx3.2$ MeV, where this function can be assumed to be applicable. Further, applying the assumption on equal contribution of positive- and negative-parity states \cite{Lokitz04,Kalmykov07}, the density of $J=1^-$ states was obtained.
In contrast to the previously published results on $^{96}$Mo \cite{Martin2017} and $^{208}$Pb \cite{Bassauer2016}, there is in fact a region of overlap between the two data sets, as shown in Fig.~\ref{fig: LD_Oslo_Darmstadt}. 
The Oslo data, as well as the CT model used in the normalization procedure (blue dashed line), lie within, but closer to the lower edge of the error band for the inelastic proton scattering data up to $\approx 10.5 $ MeV. 
This provides support of the spin-cutoff model adopted in the Oslo-method normalization.
A model predicting a higher spin-cutoff value than presented in Table~\ref{tab:table_1} would imply a wider spin distribution and, therefore, a significantly lower fraction of $J=1$ states leading to a larger discrepancy between the Oslo and the ($p,p^{\prime})$ data in the overlapping area.
Thus, we can conclude that the spin-cutoff estimate provided by Eq.~(\ref{eq:8}) is reasonable, and probably lies closer to the upper limit in the range of acceptable spin-cutoff values that would make the two experimental NLDs agree with each other.

The constant temperature regime, characterized by the pair-breaking process, continues at least up to the neutron separation energy or higher, where the temperature begins to rise and the Fermi gas behaviour of nucleons sets in. 
As shown in Fig.~\ref{fig: LD_Oslo_Darmstadt}, the CT model  begins to deviate quite drastically from the ($p,p^{\prime})$ data at higher excitation energies, well above the $S_n$ value. 
For this reason, the BSFG model is expected to provide a more accurate description of the NLD at high excitation energies, although it is not an appropriate model at lower excitation energies.
The global fit of all data with the BSFG model only indeed fails to reproduce the regime of increasing nuclear temperature between $\approx 6.5-14$ MeV, especially in the vicinity of the neutron separation energy and slightly above. 
The composite NLD formula, introduced by  Gilbert and Cameron in Ref.~\cite{Gilbert65} (denoted as G\&C), combines the CT model at lower excitation energies and the BSFG model at higher energies, and appears to be more suitable for the simultaneous description of the Oslo and ($p,p^{\prime}$) data. 
From the result of the fit with the composite NLD formula, the constant temperature regime holds up to $\approx 8.5$ MeV, \textit{i.e.} in the vicinity of the neutron separation energy. Even though this formula reproduces the general trend and performs better than the BSFG, it is still not able to completely describe the NLD above the neutron separation limit.

Microscopic spin- and parity-dependent NLD calculations based on the Hartree-Fock-Bogoliubov plus combinatorial method \cite{Goriely2008} deviate from both the Oslo and the ($p,p^{\prime})$ data throughout the whole energy range (from 3.2 to 14 MeV), being higher by a factor of $\approx$4-5 on average. 
On the other hand, the NLD calculated within the temperature-dependent Hartree-Fock-Bogolyubov approach with the Gogny force \cite{Hilaire2013} follows the ($p,p^{\prime})$ data and the composite formula prediction nicely from $\approx6.5$ MeV excitation energy and above, while still being about a factor of 3 higher than the Oslo-method NLD.
For the case of the total NLD, this deviation might reach up to two orders of magnitude.
We conclude that although microscopic models are appealing, as they should in principle grasp the underlying physics in contrast to simple analytical formulae, they are at this point not able to describe experimental data well enough over a wide excitation-energy range. 

\section{\label{sec 4: PT}Porter-Thomas fluctuations and $\gamma-$ray strength functions}

The experimental GSFs extracted with the Oslo method result from averaging $\gamma$ transitions over relatively wide excitation-energy windows, $\approx 4.6$ for $^{120}$Sn and 3.5 MeV for $^{124}$Sn (region 3 in Fig.~\ref{fig: matrices}(c) and (f)). 
Therefore, any variations of the strength due to  PT fluctuations are expected to be strongly suppressed, lying well within the estimated error bands. 
As such, PT fluctuations play a minor role and have little influence on the overall shapes of the GSFs. 
However, to test the gBA hypothesis, it is necessary to investigate how the GSF varies as a function of excitation energy (and also, in principle, spin and parity of the initial and final states). 
Then, a complication arises, because the action of narrowing down the averaging interval to study the GSF for different specific initial and final excitation energies will inevitably introduce larger uncertainties due to increased PT fluctuations of the partial radiative widths.

Oslo-method data have previously been used to study the shapes of the GSFs as functions of initial and final excitation energies to address the question on the validity of the gBA hypothesis \cite{Guttormsen2011_BA, Guttormsen2016, Guttormsen2016_2, Campo2018}. 
With the exception of Ref.~\cite{Campo2018}, which presents a detailed discussion and estimates of the PT fluctuations for the case of $^{64,65}$Ni, the role of these fluctuations are approached mostly in a qualitative way. 
Due to the particularly high density of initial and accessible final states in $^{238}$Np, studied in Ref. \cite{Guttormsen2016}, reaching up to $\approx 4.3\cdot10^6$ states at $S_n=5.488$ MeV, the PT fluctuations are expected to be negligible for the comparison of individual GSFs for different individual initial and final excitation energies with the Oslo-method strength. An excellent agreement of all strengths was found, and this indeed serves as a strong argument for the validity of the gBA hypothesis \cite{Guttormsen2016}. 
Such a comparison, however, is much more difficult in the case of lighter nuclei such as $^{46}$Ti~\cite{Guttormsen2011_BA}, $^{64,65}$Ni~\cite{Campo2018}, and $^{92}$Zr~\cite{Wiedeking2020}. 
For example, the density of levels at $S_n=9.658$ MeV in $^{64}$Ni is only $\approx 2.6\cdot10^3$ MeV$^{-1}$, and variations on the strengths for specific excitation energies might reach some tens of percent of the absolute value \cite{Campo2018}.
In this regard, the nuclei studied in this work present an intermediate case between the heavy $^{238}$Np and relatively light $^{64,65}$Ni nuclei, with the total NLDs of $\approx 2.5\cdot10^5$ MeV$^{-1}$ at $S_n=9.104$ MeV for $^{120}$Sn and $\approx 8.8\cdot10^4$ MeV$^{-1}$ at $S_n=8.489$ MeV for $^{124}$Sn.

To study the variation in the GSFs of $^{120,124}$Sn, we follow the procedure outlined in Refs.~\cite{Guttormsen2016_2, Campo2018}, assuming that the fluctuations of the GSF follow a $\chi^2_{\nu}$ distribution with the number of degrees of freedom corresponding to the number of $\gamma$-ray transitions $n(E_{\gamma})$ at a given transition energy $E_{\gamma}$. 
Relative fluctuations of the GSF are given by the ratio between the deviation $\sigma_{PT}$ and average $\mu$, or $r=\sigma_{PT}/\mu=\sqrt{2/\nu}$, of the $\chi^2_{\nu}$ distribution \cite{Porter1956}.

The number of transitions (\textit{i.e.}, the number of partial widths, or primary transitions) $n$ can be calculated for each $E_{\gamma}$  for specific initial and final excitation energies, allowing to study how the fluctuations evolve with $\gamma$-ray and excitation energy. 
We adopt the following relation from Refs.~\cite{Guttormsen2016_2, Campo2018} to estimate the number of transitions $n(E_{\gamma},E_i)$:
\begin{equation}
   \label{eq:15} 
   \begin{split}
   n(E_{\gamma}, E_i)=&\Delta E^2\sum_{J\pi}\sum_{L=-1}^1\sum_{\pi^{\prime}}\rho(E_i, J, \pi)\times\\
   &\times\rho(E_i-E_{\gamma}, J+L, \pi^{\prime}),
   \end{split}
\end{equation}
where we consider dipole transitions only, and $\Delta E$ is the excitation energy bin width. By substituting $E_i$ with $E_f$ and $E_i-E_{\gamma}$ with $E_f+E_{\gamma}$, it is also possible to obtain the number of transitions as a function of $E_{\gamma}$ and final excitation energy.

We limit ourselves to two types of cases in estimating the GSF fluctuations. 
Firstly, we study the case when the initial $E_i$ and final $E_f$ excitation energies both lie within the quasi-continuum region, for which the spin distribution of Eq.~(\ref{eq:6}) is considered applicable. 
This allows to apply this distribution to account for the spin dependence of the NLDs in Eq.~(\ref{eq:15}). Further, it is assumed again an equal contribution of positive- and negative-parity states within the quasi-continuum. 
We also require a minimum level density of 10 levels per bin, corresponding to $E_f\approx 3.2$ MeV in $^{120}$Sn and $E_f\approx 3.0$ MeV in $^{124}$Sn.
Note that this is a rather crude estimate that should be taken with some caution. 
However, since we want to obtain an approximate magnitude of the fluctuations, small deviations from the spin distribution formula are not expected to impact the results. 
Secondly, we consider initial excited states within the quasi-continuum and final states with known parities and spins within the discrete region. 
Here, the level density at the final excitation energy can be calculated directly using the known states from Ref.~\cite{ensdf}. 

\begin{figure}[t]
\includegraphics[width=1.0\columnwidth]{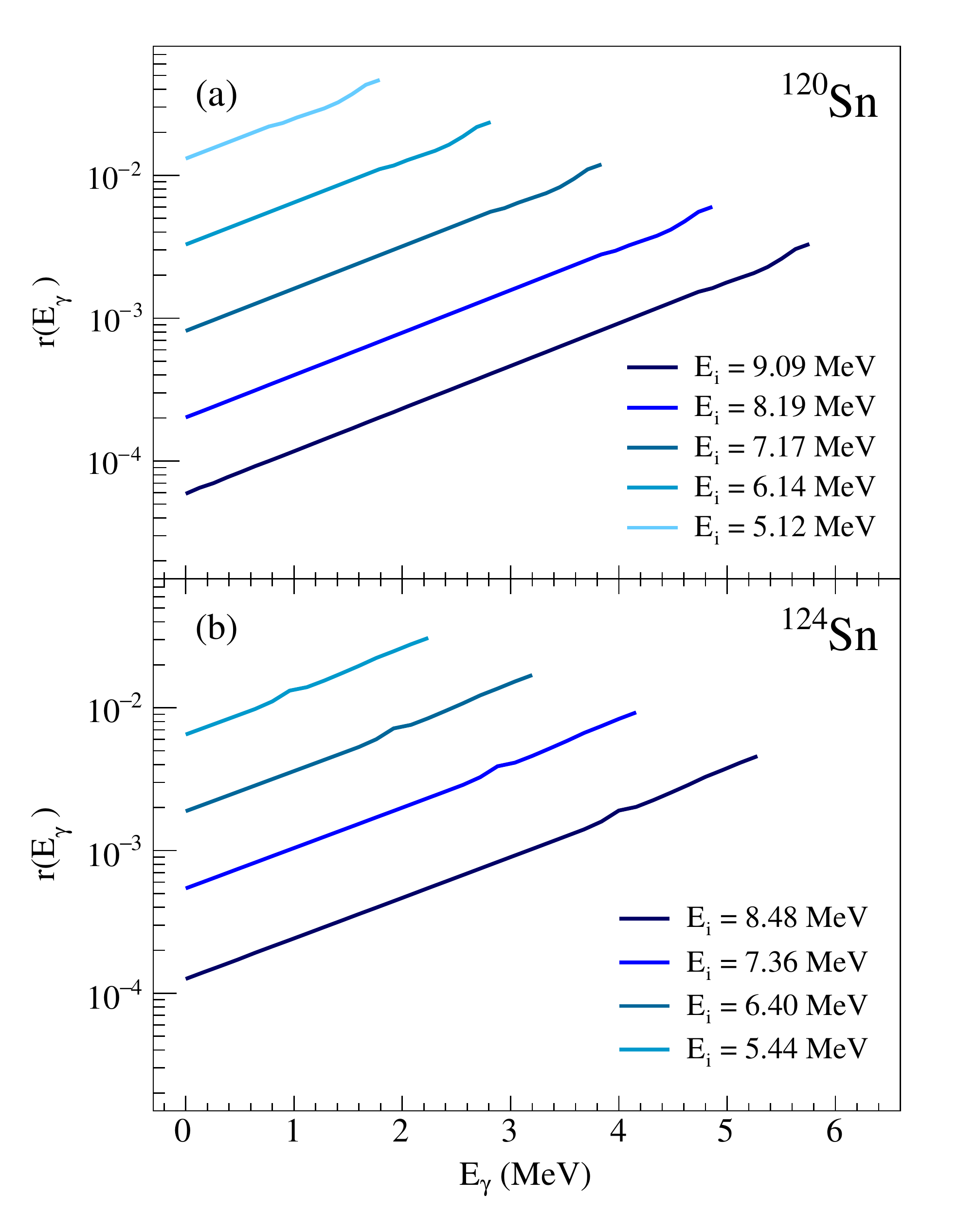}
\caption{\label{fig: PT init}
Relative fluctuations of the GSF $r(E_{\gamma}, E_i)$ for different initial excitation energies for (a) $^{120}$Sn and (b) $^{124}$Sn. All initial $E_i$ and final energies $E_i-E_{\gamma}$ lie within the quasi-continuum region. The excitation and $\gamma$-ray energy bins are 128 keV for $^{120}$Sn and 160 keV for $^{124}$Sn.    
}
\end{figure}

\begin{figure}[t]
\includegraphics[width=1.0\columnwidth]{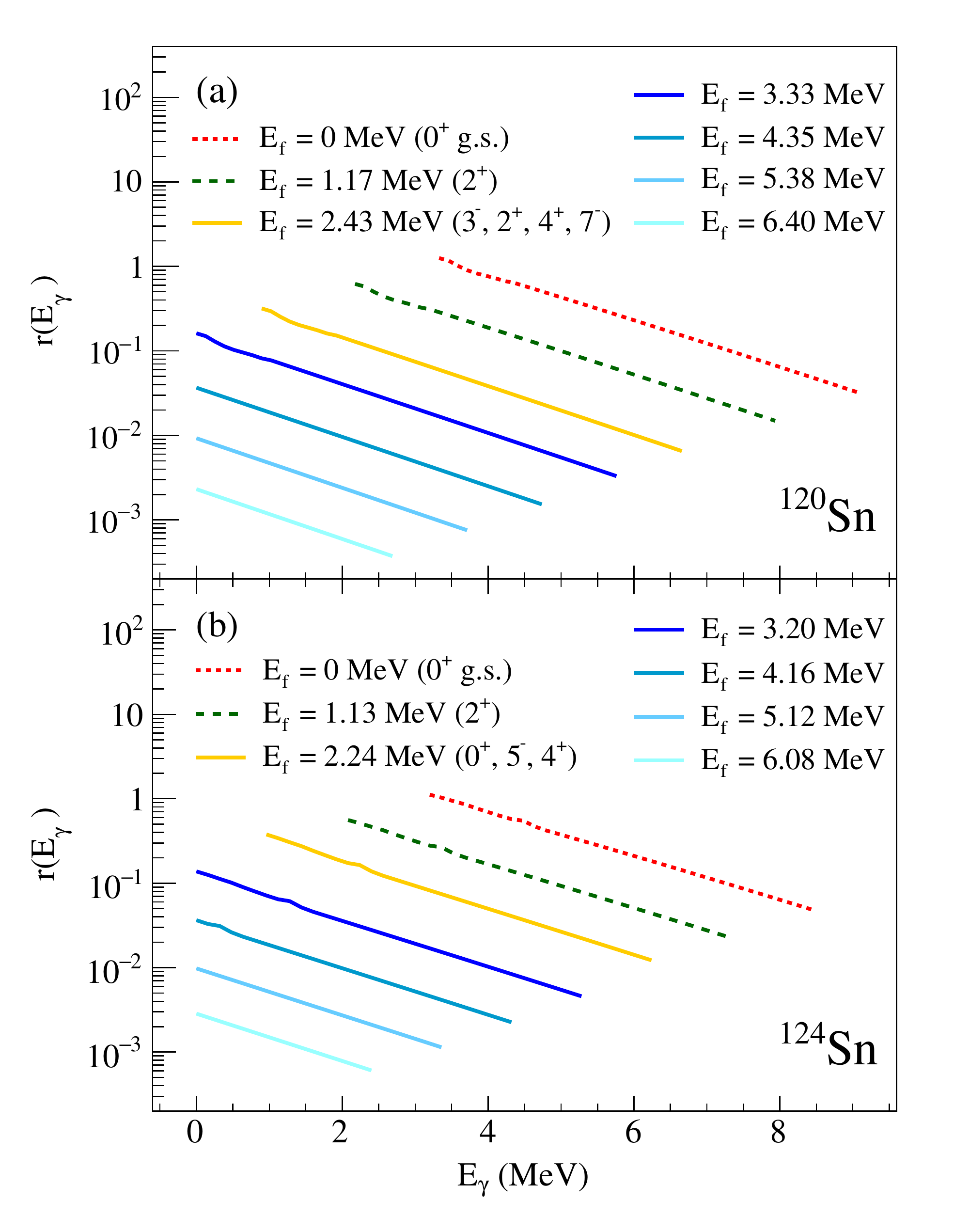}
\caption{\label{fig: PT fin}
Relative fluctuations of the GSF $r(E_{\gamma}, E_f)$ for different final excitation energies for (a) $^{120}$Sn and (b) $^{124}$Sn. All initial energies $E_i-E_{\gamma}$ lie within the quasi-continuum region.  The same applies to the different final energies $E_f$ represented by blue lines. The red dashed line corresponds to the ground state as the final state, the green one corresponds to the first excited $2^+$ state as the final state, and the yellow one corresponds to several discrete final low-lying states. The excitation and $\gamma$-ray energy bins  are 128 keV for $^{120}$Sn and 160 keV for $^{124}$Sn.     
}
\end{figure}

Figure \ref{fig: PT init} shows the relative GSF fluctuations $r(E_{\gamma}, E_i)=\sqrt{2/n(E_{\gamma}, E_i)}$ as functions of $E_{\gamma}$ for transitions from different initial excitation-energy bins within the quasi-continuum for $^{120}$Sn and $^{124}$Sn. 
The data are shown for $E_f\geq3.2$ MeV for $^{120}$Sn and  $E_f\geq3.0$ MeV for $^{124}$Sn, so that the final excitation energies of the included transitions lie within the quasi-continuum. 
The experimental level densities were used for the calculation. 
Similar to the results for $^{64,65}$Ni \cite{Campo2018}, the fluctuations increase exponentially with $\gamma$-ray energy for a given $E_i$, as well as from the lowest to the highest initial excitation energy at a given $E_{\gamma}$. 
This behaviour can  easily be explained by the decreasing number of possible transitions for consecutively lower excitation energies, given the exponentially decreasing density of accessible levels.

The magnitudes of the fluctuations in both nuclei are quite similar due to the similar values of the total NLDs, and all minor differences stem primarily from a slight difference in the bin width. 
At the neutron separation energy, fluctuations in both nuclei range from  $\approx 10^{-4}$ to $4-5\cdot 10^{-3}$ \%, while for the lower excitation energy they reach up to $\approx 3-6\%$. 
Fluctuations of these orders of magnitude are indeed expected for the relatively heavy $^{120,124}$Sn nuclei. 
For example, based on the NLD of $^{64}$Ni \cite{Campo2018} and $^{120}$Sn, the number of transitions at $E_i\approx 7.7$ MeV at $E_{\gamma}\approx2.3$ MeV in $^{120}$Sn is roughly by a factor of 1000 larger than in $^{64}$Ni, which indeed yields larger fluctuations in $^{64}$Ni by approximately a factor of 30. 

The relative GSF fluctuations calculated from the transitions to specific final excitation energies demonstrate an opposite trend, exponentially decreasing with $\gamma$-ray energies, as shown in Fig.~\ref{fig: PT fin}. 
These trends are displayed with an approximately equal spacing for several final excitation energy bins within the quasi-continuum, as well as the bins containing the ground state, the first excited state, and several known low-lying excited states. 
In contrast to the lowest initial excitation energies, fluctuations at final excitation energies below $E_f\approx 3$ MeV reach up to tens of percent and might become a considerable contribution to the total uncertainty of the GSF. 

\begin{figure*}[t]
\includegraphics[width=1.0\textwidth]{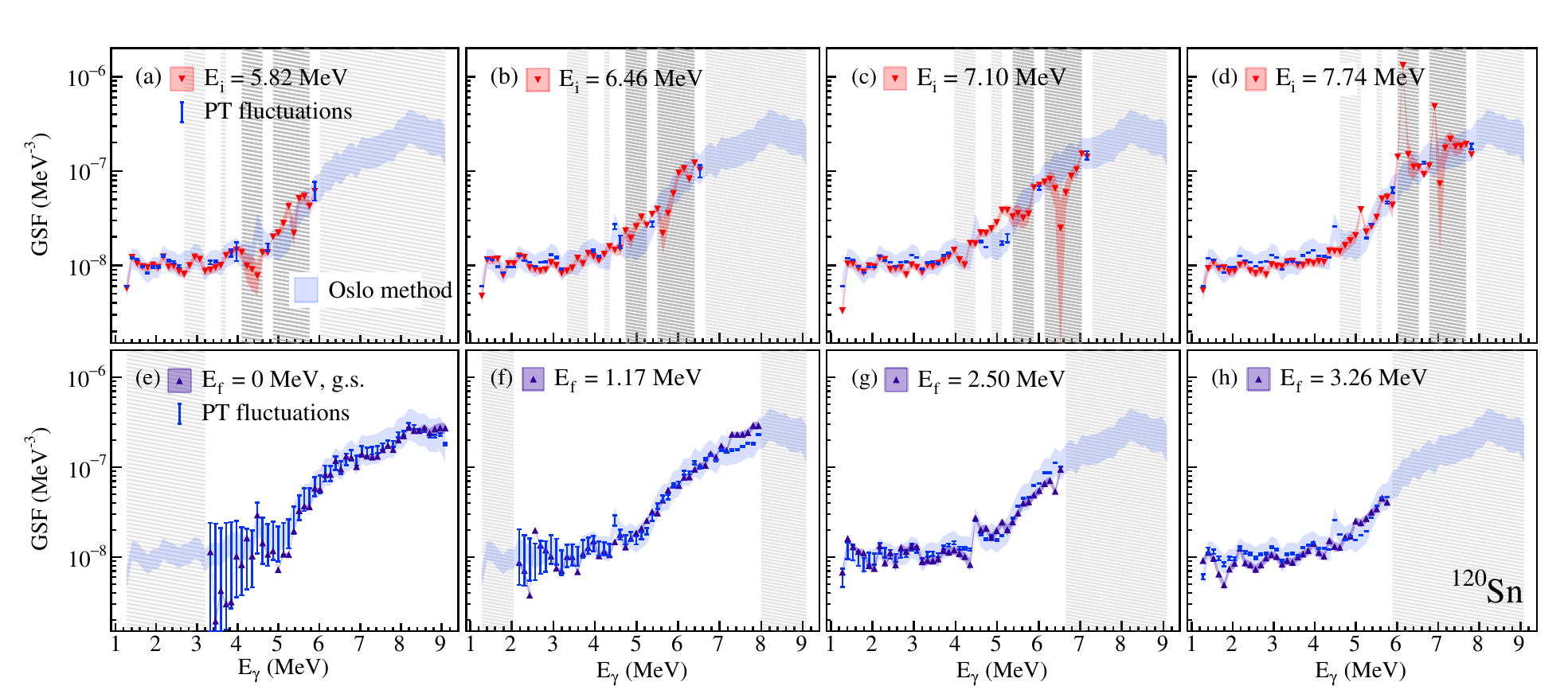}
\caption{\label{fig: Radex 120Sn} GSFs for $^{120}$Sn at initial excitation energies (a) 5.82 MeV, (b) 6.46 MeV, (c) 7.10 MeV, (d) 7.74 MeV and final excitation energies (e) ground state, (f) first excited state, (g) 2.50 MeV, (h) 3.26 MeV compared to the Oslo method strength (blue shaded band). For each strength the statistical error band is shown together with the error due to the PT fluctuations. Dark grey regions correspond to the areas of expected infinite PT fluctuations, light grey area marks energies for which the fluctuations of the strength were not determined. The $\gamma$-ray and excitation energy bin widths are both 128 keV.
}
\end{figure*}
\begin{figure*}[t]
\includegraphics[width=1.0\textwidth]{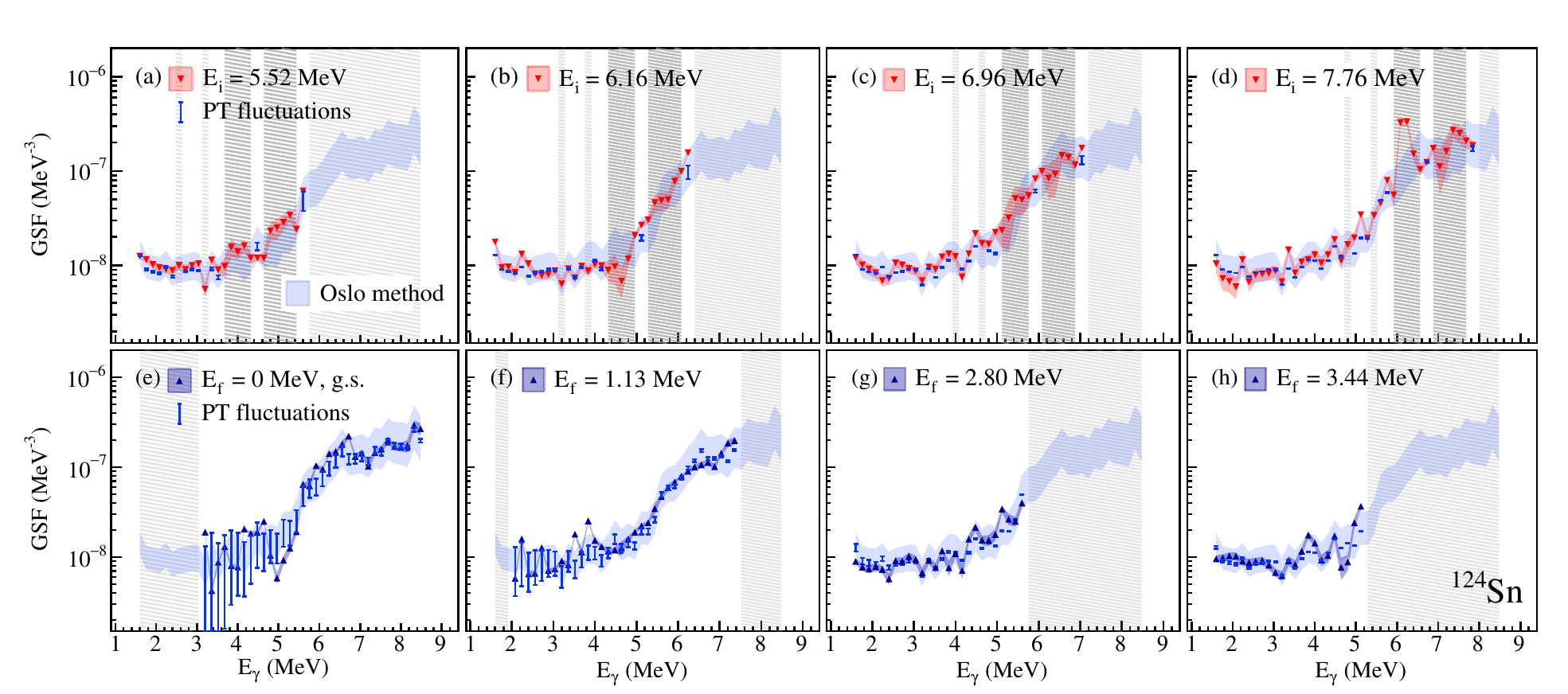}
\caption{\label{fig: Radex 124Sn} GSFs for $^{124}$Sn at initial excitation energies (a) 5.52 MeV, (b) 6.16 MeV, (c) 6.96 MeV, (d) 7.76 MeV and final excitation energies (e) ground state, (f) first excited state, (g) 2.80 MeV, (h) 3.44 MeV compared to the Oslo method strength (blue shaded band). For each strength the statistical error band is shown together with the error due to the PT fluctuations. Dark grey regions correspond to the areas of expected infinite PT fluctuations, light grey area marks energies for which the fluctuations of the strength were not determined. The $\gamma$-ray and excitation energy bin widths are both 160 keV.
}
\end{figure*}

The estimates of the PT fluctuations can be further put into the context of testing the gBA hypothesis for $^{120,124}$Sn. 
By analogy with the $^{238}$Np results from Ref.~\cite{Guttormsen2016}, the experimental data obtained for $^{120,124}$Sn can be readily used to test whether the transmission coefficients, and, therefore, the GSFs, are dependent on the initial and final excitation energies. 
Equation~(\ref{eq:2}) can be rewritten in the form \cite{Guttormsen2016}:
\begin{equation}
    \label{eq:16}
    P(E_{\gamma},E_i)N(E_i) = \mathcal{T}(E_{\gamma}) \cdot\rho(E_i-E_{\gamma}),
\end{equation}
where we introduce an additional energy-dependent factor $N(E_i)$ given by:
\begin{equation}
    \label{eq:17}    
N(E_i)=\frac{\int_0^{E_i}\mathcal{T}(E_{\gamma}) \cdot\rho(E_i-E_{\gamma})dE_{\gamma} }{\int_0^{E_i}P(E_{\gamma},E_i)dE_{\gamma}}.
\end{equation}
Here, we make use of the transmission coefficient extracted from the Oslo method, and hence averaged over a wide range of excitation energies. 
We can deduce the transmission coefficient as a function of excitation energy and $\gamma$ energy through
\begin{equation}
    \label{eq:18}  
    \mathcal{T}(E_{\gamma}, E_i) = \frac{P(E_{\gamma},E_i)N(E_i)}{\rho(E_i-E_{\gamma})}.
\end{equation}
A similar relation can be obtained for the final excitation energy by substituting $E_i$ with $E_f+E_{\gamma}$.

The GSFs for several initial excitation energies in the case of $^{120}$Sn were previously published in Ref.~\cite{Markova2021}, where they were compared with the strength extracted with the Oslo method. 
In this work, we present the comparison of the individual GSFs for different initial and final excitation energy bins for both $^{120}$Sn and $^{124}$Sn with the corresponding Oslo-method results. 
Individual strengths are shown together with the error band due to the statistical uncertainty propagated through the method, denoted by statistical for short. 
As the Oslo-method GSF is an averaged strength with heavily suppressed PT fluctuations, it is shown with the total error band as well as additional error bars, denoting the expected  PT fluctuations, or rather expected deviations of the individual strengths due to PT fluctuations. 
The latter is essential to assess whether there is an agreement or not  between the strengths extracted for various excitation-energy bins and the Oslo-method strengths.

The results for $^{120}$Sn at four initial excitation energies are shown in the upper row of Fig.~\ref{fig: Radex 120Sn}.
The dark grey shaded areas indicate regions of potential infinite fluctuations due to the expected zero values of the NLD at the final excitation energies in the energy gaps between the first few discrete states. 
As  can be seen from Fig.~\ref{fig: LD}, the experimental NLD has small non-zero values between the ground state and the first and second excited states at $\approx 1.171$ and $1.875$ MeV due to the experimental resolution and the presence of some residual counts in the raw matrix after the background subtraction.
The analysis applied to each individual excitation energy $E_i$ generates a continuous data set for the GSF from the highest possible gamma-ray energy at $E_{\gamma}=E_i$ downward to gamma-ray energies below 2 MeV shown for $^{120}$Sn in Fig.~\ref{fig: Radex 120Sn}(a)-(d). The GSF values in the dark grey region at higher gamma energies belong to hypothetical primary gamma-ray transitions in the energy range between the ground state and 1.171 MeV, while the dark grey region at lower energies belongs to decays into the energy range from 1.171 MeV to 1.875 MeV. However, it should be mentioned that direct gamma decays to those final excitation energy regions are physically not possible and that the corresponding data points are artifacts of the continuous analysis. It is, however, interesting to observe that the PT fluctuation analysis reveals those regions by unusually large PT fluctuations

In case of fixed initial excitation energies, light grey shaded areas correspond to energy bins where the fluctuations can not be estimated either due to to $E_{\gamma}>E_i$ or unambiguous spins of some final excited states. In the latter case it is no longer possible to define what spins of initial states within the quasi-continuum yielding dipole transitions must be included to the sum in Eq.~\ref{eq:15} . For the rest of the experimental points, the fluctuations were estimated and shown in Fig.~\ref{fig: Radex 120Sn} as vertical error bars. The values of these errors exceed or are of the same magnitude as the statistical uncertainties for high $E_{\gamma}$ for all of the presented cases. 
For the highest initial excitation energies in Fig.~\ref{fig: Radex 120Sn}, $E_i=7.74$ and 7.10 MeV, they become increasingly suppressed as compared to the statistical errors, by roughly a factor of $10$ at $E_{\gamma}\approx4.5$ MeV, gradually increasing to $\approx10^2$ toward $E_{\gamma}\approx 1$ MeV. 
For lower initial excitation energies, this factors are of  order 1 and 10. 
Except for the strong deviations in the areas with expected large fluctuations (dark grey areas), all strengths are in fairly good agreement with the Oslo-method result within its error band.

Similar results with an excitation energy bin width of 160 keV are shown for $^{124}$Sn in the upper row of Fig.~\ref{fig: Radex 124Sn}. 
Since the range of populated spins might be limited in this case, using the total NLD provides a lower estimate of the PT fluctuations, and they might be slightly larger than shown in the figure. 
By analogy with the case of $^{120}$Sn, the GSFs for different initial excitation energies are in rather good agreement with the Oslo-method strength within the shown error bands and areas of expected finite PT fluctuations.  
These results for both the $^{120,124}$Sn isotopes bring further support to the GSF being independent on the initial excitation energy, in accordance with the gBA hypothesis.

\begin{figure}[t]
\includegraphics[width=1.0\columnwidth]{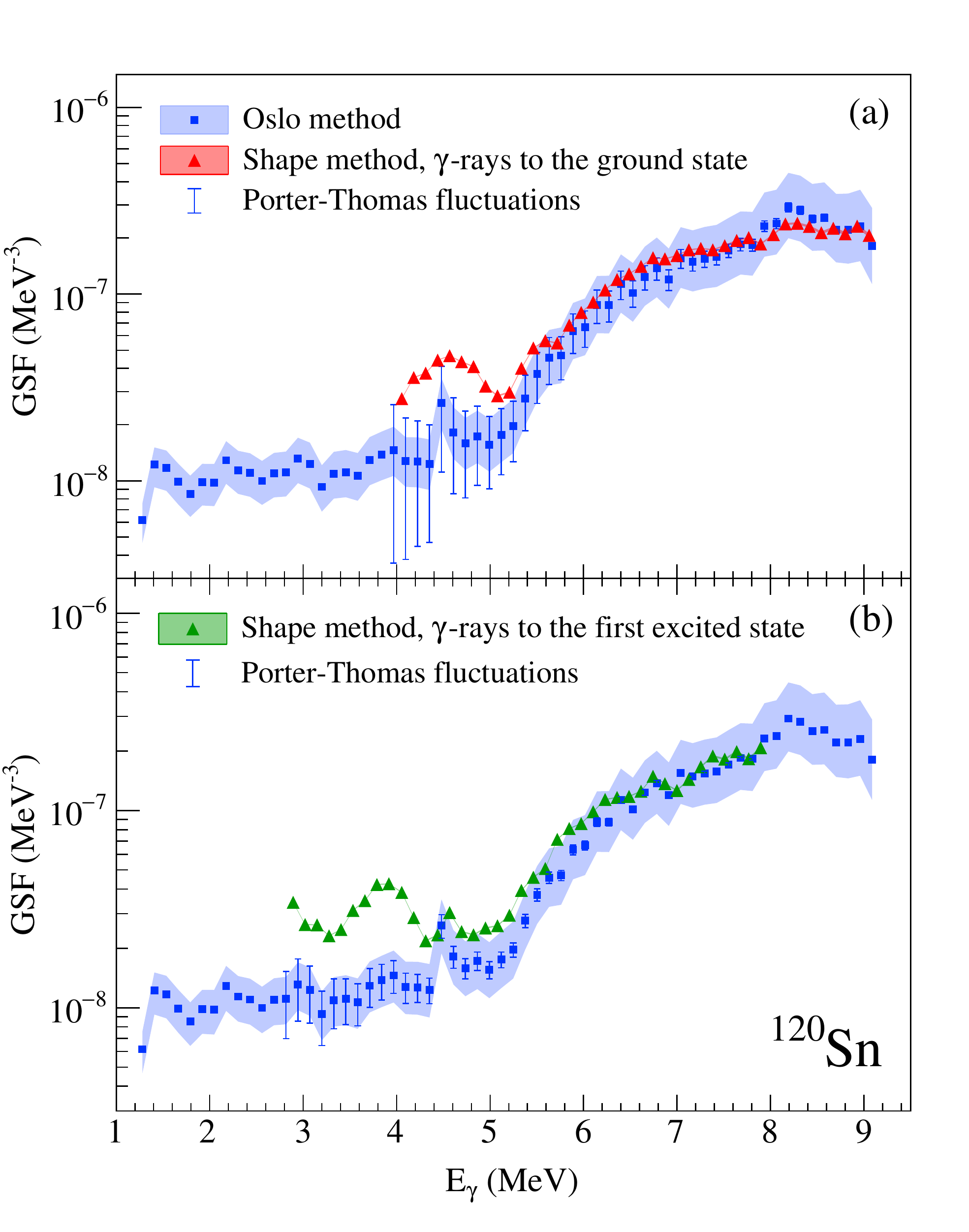}
\caption{\label{fig: SM 120Sn}
Shape-method GSFs of $^{120}$Sn for $\gamma-$rays feeding the ground state (a) and the first excited state (b)  compared to the Oslo method result (blue band). The Shape method results are shown together with the statistical error propagated through the method, shown as a band (significantly smaller in width than the size of the data points), and the error bars due to the PT fluctuations. The Oslo method GSF is shown with the total (stat.+syst.) error band.
}
\end{figure}
\begin{figure}[t]
\includegraphics[width=1.0\columnwidth]{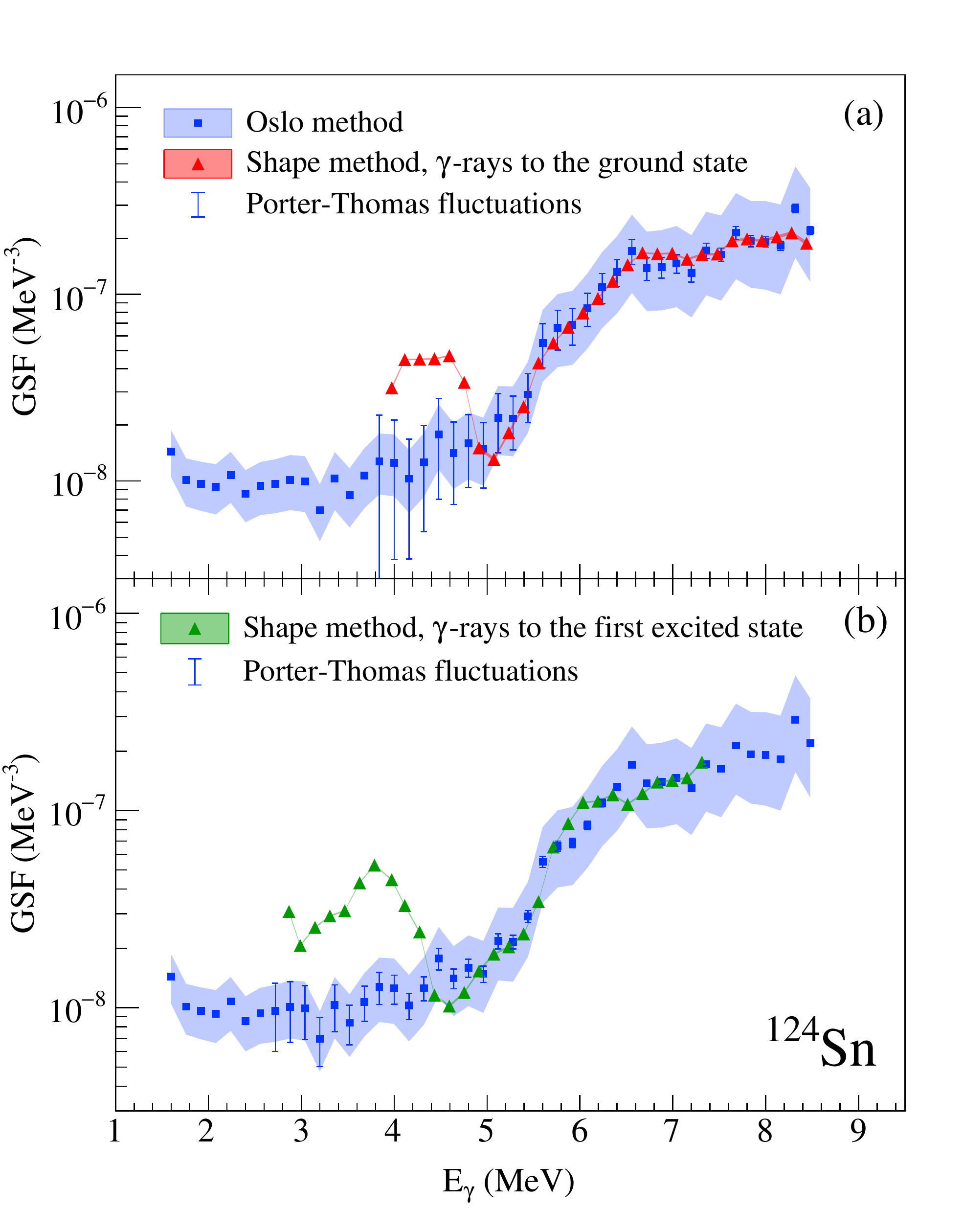}
\caption{\label{fig: SM 124Sn}
Same as Fig.~\ref{fig: SM 120Sn}, but for $^{124}$Sn.
}
\end{figure}
At lower excitation energies, the uncertainty due to  PT fluctuations is expected to gradually outweigh the statistical error bar. 
This effect becomes especially apparent for the GSFs extracted for specific final excitation-energy bins. 
The GSF for the ground state and the first excited state at $1.171$ MeV in $^{120}$Sn are demonstrated in comparison with the Oslo-method GSF in Figs. \ref{fig: Radex 120Sn}(e) and (f). 
The data are shown for $E_f+E_{\gamma}\geq 3.2$ MeV. The area below this energy and the area corresponding to $E_f+E_{\gamma}>S_n$ are shaded. The fluctuations of the ground-state strength are  large below $\approx 5$ MeV, where they reach $\approx40\%$ of the absolute value. 
Between $E_{\gamma}\approx3.3$ and 5 MeV, the fluctuations of the strength are $\approx 60\%$ on average and reach up to 90\% toward the lowest $\gamma$ energy. 
The latter case corresponds to only 1-3 possible dipole transitions at this $E_{\gamma}$.
Applying  the $\chi^2_{\nu}$ distribution for  fluctuations of so few transition widths is not justified as it is valid solely in the statistical regime. 
Thus, the estimation procedure should be taken with great care when $r(E_{\gamma})$ approaches values of 1. 

Below $\approx 5$ MeV, some strong deviations of the ground state strength from the Oslo-method result are observed.
Besides the strong PT fluctuations at these $\gamma$-ray energies, there might be some quadrupole transitions that cause methodical problems in this region. 
As the extraction of the GSF relies on dipole radiation being dominant, quadrupole transitions from numerous low-lying $2^+$ states to the ground state could distort the strength as the  factor of $E_{\gamma}^{5}$ should be used instead of $E_{\gamma}^{3}$. 
At higher $\gamma$-ray energies, the ground-state strength reproduces the slope of the Oslo method strength, lying well within the Oslo-method error band. 
Similar effects can be seen for the $^{124}$Sn (in Fig.~\ref{fig: Radex 124Sn}(e), $E_f+E_{\gamma}\geq 3.0$ MeV), where the fluctuations were again estimated with the total NLD and, therefore, should be considered lower-limit estimates.

High PT fluctuations of 10-60\% are observed also for the GSF to the first excited states in both isotopes, as shown in Fig.~\ref{fig: Radex 120Sn}(f) and \ref{fig: Radex 124Sn}(f). For both nuclei these strengths reproduce the slopes of the Oslo method GSF in the region between 5 and 6.5 MeV quite well. For the higher final excitation energies, the fluctuations of the strengths are at most by one order of magnitude larger than the statistical uncertainties at low $\gamma-$ray energies, whilst at higher $\gamma-$ray energies they are by one order of magnitude smaller. For these strengths it is challenging to argue for an \textit{exact} agreement with the Oslo method result. If taking a general agreement of the strengths within the error bars as a criterion, it can be possible to claim an overall independence of the strengths of final excitation energy for $^{120,124}$Sn.

As the PT fluctuations become more significant at lower final excitation energies, they are expected to make a considerable contribution to the total error band of the Shape-method results. 
In figures \ref{fig: SM 120Sn}  and \ref{fig: SM 124Sn}, the GSFs for $\gamma$ rays feeding the ground state and the first excited $2^+$ state are shown for $^{120}$Sn and $^{124}$Sn, respectively, together with the corresponding Oslo-method strengths. 
To test what a reasonable minimum excitation-energy limit would be for the application of the Shape method, we choose $E_i =4$ MeV in both nuclei as a starting point.
The Shape method results are presented with their statistical uncertainties, propagated through the unfolding and the first generation method. 
The Oslo-method strength is shown with the total error band and the expected variations of the corresponding ground-state or first-excited state strengths due to the PT fluctuations. 
Both of these strengths for $^{120}$Sn follow the shape of the Oslo-method strength quite well from the neutron separation energy and down to $\approx 5.5-6$ MeV. 
Here, they  start deviating gradually for lower $\gamma$-ray energies. 
In $^{124}$Sn, the agreement between the GSFs is quite good  from $E_\gamma \approx 5$ MeV and higher.

Remarkably, the ground-state strengths and the first-excited state strengths for $^{120,124}$Sn demonstrate quite significant enhancements between 3 and 5 MeV, which can not be attributed to any real features of the strength.
Moreover, there are no noticeable structures on the diagonals at $4< E_i< 5$ MeV that might have induced these features. 
No similar effect was previously reported for even-even isotopes \cite{Wiedeking2020}. 
The appearance of these bumps might partly arise from the failure of the internal normalisation technique at relatively low $\gamma$-ray energies where large fluctuations of the strengths are observed.
The fluctuations of the ground-state strength in $^{120,124}$Sn range from $\approx$30 to 70\% below 5.5 MeV, and from $\approx$15-35\% below 4.3 MeV for the GSF corresponding to the first excited state. 
Since the pairs of data points for the two diagonals at each excitation energy are normalized internally to each other (see Ref.~\cite{Wiedeking2020}), large variations of the strengths could lead to an erratic internal normalization at relatively low $\gamma$-ray energies. 
When reaching densities of 1$\cdot10^3$-2$\cdot10^3$ levels per MeV, the distorting effect due to the PT fluctuations becomes smaller, and the Shape-method results follow nicely the Oslo-method strength in both cases. 
This potential problem should be considered in future studies performed with the Shape method. 
When approaching the neutron separation energies in $^{120,124}$Sn, fluctuations of the strengths do not exceed a few percent, which is comparable to the statistical error bands shown in Figs.~\ref{fig: SM 120Sn} and \ref{fig: SM 124Sn}, whilst for the rest of the energy range, the PT fluctuations make  a noticeable contribution to the uncertainties. 

Additional explanations for the smooth bump-like structures observed in the GSF might come from the failure of some basic assumptions in the Shape method such as a symmetric parity distribution of the initial nuclear levels, pure dipole transitions of the involved $\gamma$-ray decays, and a spin-independent excitation probability in the ($p,p^{\prime}\gamma$) reaction at 16 MeV. The lower the excitation energy, the less the assumption of a symmetric parity distribution might be justified, especially in the magic Sn isotopes, so this may lead to deviations when using the Shape method at excitation energies below 5-6 MeV. Furthermore, similar to the discussion of the Oslo method, potential contributions of quadrupole transitions can distort the analysis procedure due to the different energy factor of $E_{\gamma}^5$ as compared to $E_{\gamma}^3$ for dipole transitions. In particular, the excited $2^+$ states will most likely decay (on average) preferably to the first $2^+$ instead to the ground state. Within the Shape method, this can lead to the fact that the value of the GSF for the ground state $\gamma$-decay is (on average) smaller than for the decay into the first $2^+$ state. Thus the value pair in the Shape method has an increasing course towards low gamma energies due to $f_{\text{to }2^+}[E_i-E_x(2^+)] > f_{\text{to }g.s.}[E_i-E_x(g.s.)]$  and might explain the increasing bump-like trend of the GSF. It remains an open question as to why the deviation of the strengths is systematically upward (always an increase) and whether the PT fluctuations, asymmetric parity distributions or the specific decay behavior of $2^+$ states at low excitation energies are the main cause of the observed deviation

\section{\label{sec 6: Conclusion}Conclusions}

The nuclear level densities and $\gamma$-ray strength functions of $^{120,124}$Sn were extracted using the Oslo method, and the slopes of the strengths were additionally constrained with the Shape method. 
The NLDs were found to be in good agreement with previously deduced NLDs for $^{116,118,122}$Sn, with slight deviations primarily due to some differences in the normalization procedures. 
The Oslo-method NLD for $1^{-}$ states in $^{124}$Sn is in fairly good agreement within the estimated error bands with the result obtained from the fluctuation analysis of high-resolution inelastic proton scattering spectra above 6 MeV. Given the model-independence of the ($p,p^{\prime}$) result, this agreement supports the choice of the spin distribution function and the spin-cutoff parameter employed in the Oslo method. The combined results covering excitation energies up to 14 MeV clearly demonstrate the transition between the constant temperature and the Fermi gas regimes at $\approx 7$ MeV. 

The experimental NLDs were used to estimate the role of the Porter-Thomas fluctuations in assessing the generalised Brink-Axel hypothesis below the neutron separation energy in $^{120,124}$Sn, as well as the applicability of the Shape method. 
Most of the deviations of the GSFs for different initial and final excitation energies from the Oslo-method strength can be explained by strong PT fluctuations due to very few $\gamma$ transitions. 
For the ground-state and the first-excited state strengths, this effect is especially apparent, with the PT fluctuations reaching up to 90-100\%  at low $\gamma$-ray energies. 
Despite some local discrepancies, the individual GSFs are in overall good agreement with the Oslo-method strength within the error bands, suggesting an independence of initial and final excitation energies in support of the generalized Brink-Axel hypothesis within uncertainties of the Oslo method. 

Strong PT fluctuations were found to play a noticeable role in the extraction of the GSFs with the Shape method, as they might contribute to considerable deviations from the Oslo-method result at low $\gamma$-ray energies. 
The reliability of the Shape method applied to $^{120,124}$Sn is under question for values of the NLDs below 1$\cdot10^3$-2$\cdot10^3$ levels per MeV, but quite satisfactory above this limit in both nuclei. 
Further investigations are needed to understand why the Shape method seemingly leads to an overestimate of the low-energy strength in the region where the PT fluctuations are large.

\begin{acknowledgments}
The authors express their thanks to J.~C.~M\"{u}ller, P.~A.~Sobas, and J.~C.~Wikne at the Oslo Cyclotron Laboratory for operating the cyclotron and providing excellent experimental conditions.
A.~Zilges is sincerely thanked for stimulating discussions and for providing the $^{120, 124}$Sn targets.
This work was supported in part by the National Science Foundation under Grant No.\ OISE-1927130 (IReNA), by the Deutsche Forschungsgemeinschaft (DFG, German Research Foundation) under Grant No.\ SFB 1245 (project ID 279384907), by the Norwegian Research Council Grant 263030, and by the National Research Foundation of South Africa (Grant No.\ 118846).
A.~C.~L. gratefully acknowledges funding by the European Research Council through ERC-STG-2014 under Grant Agreement No.\ 637686, 
and from the Research Council of Norway, project number 316116. J. I. acknowledges the support by the State of Hesse within the Research Cluster ELEMENTS (Project ID 500/10.006) and within the LOEWE program “Nuclear Photonics"

\end{acknowledgments}


\bibliographystyle{plain}
\bibliography{tin_2021}
\end{document}